\documentclass[prd,aps,nofootinbib,floatfix,11pt]{revtex4}
\usepackage{amsmath,graphicx,epsfig,amssymb,dsfont,mathtools,}
\usepackage[usenames]{color}
\usepackage{ulem} 
\usepackage{bigstrut}
\usepackage{slashed}
\usepackage{multirow}
\usepackage{subfigure}

\allowdisplaybreaks

\begin{document}\title{The study of doubly charmed pentaquark  $c c \bar qqq$ with the SU(3) symmetry}
\author{
   Ye Xing~$^{1}$~\footnote{Email:xingye\_guang@cumt.edu.cn},
   Yuekun Niu~$^{1}$~\footnote{Email:nyk@cumt.edu.cn}}

\affiliation{$^{1}$  School of Physics, China University of Mining and Technology, Xuzhou 221000, China\\
  }

\begin{abstract}
We study the masses and lifetimes of doubly charmed pentaquark $P_{cc\bar qqq}(q=u,d,s)$ primarily. The operation of masses carried out by the doubly heavy triquark-diquark model, whose results suggests the existence of stable states $cc\bar s ud$ with the parity $J^P=\frac{1}{2}^-$. The roughly calculation about lifetimes show the short magnitudes, $(4.65^{+0.71}_{-0.55})\times 10^{-13}s $ for the parity $J^P=\frac{1}{2}^-$ and $(0.93^{+0.14}_{-0.11})\times 10^{-12}  s $ for $J^P=\frac{3}{2}^-$.
Since the pentaquark $cc\bar s ud$ is interpreted as the stable bound states against strong decays, then we will focus on the production and possible decay channels of the pentaquark in the next step, the study would be fairly valuable supports for future experiments. For completeness, we systematically studied the production from $\Omega_{ccc}$ and the decay modes in the framework SU(3) flavor symmetry, including the processes of semi-leptonic and two body non-leptonic decays. Synthetically, we make a collection of the golden channels.
\end{abstract}
\maketitle

\section{Introduction}
In 2015, the LHCb collaboration announced the findings of first pentaquark states $P_c(4380)^+$ and $P_c(4450)^+$ in the decay $\Lambda_b^0\to J/\psi pK^-$, the masses and decay widths were measured respectively, $M(P_c(4380))=(4380\pm 8\pm 29)$ MeV, $\Gamma=(205\pm18\pm86)$ MeV and $M(P_c(4450))=(4449.8\pm 1.7\pm 2.5)$ MeV, $\Gamma=(39\pm5\pm19)$ MeV~\cite{Aaij:2015tga}. Subsequently, in 2019, the LHCb collaboration reported a new pentaquark $P_c(4312)^+$~\cite{Aaij:2016phn}, in addition, the analyses revealed that $P_c(4450)^+$ observed previously actually were the average of two narrow resonances $P_c(4440)^+$ and $P_c(4457)^+$. It is far from clear about the pentaquark states currently, regardless of their internal structures or production and decay behaviors. However, many theoretical studies devoted to the pentaquark have achieved remarkable results, for instance, effective field theory discuss the possible configurations of pentaquark~\cite{Liu:2019tjn,He:2019ify,Meng:2019ilv}, QCD sum rules study the masses of $P_c(4380)$ pentaquark~\cite{Azizi:2018bdv,Chen:2015moa}, and quark model study the masses~\cite{Richard:2019fms,Yang:2015bmv,Huang:2015uda}, etc.
At present, the discovered pentaquarks are largely what is called hidden charmed pentaquark with charmed quarks pair $c\bar c$, the states with two heavy charmed quarks $P(cc\bar qqq)$ are similarly available in quark model, even now, indefinitely for the existence, controversially for the stability and unclearly for the properties of the doubly charmed pentaquark states, in turn, which exactly offer the possible for the studying of strong force in greater detail and the understanding of quantum chromodynamics. Besides, the two charmed quarks in pentaquark mean more scales, which might be an ideal probe to study the interplay between perturbative QCD and non-perturbative QCD, eventually leading the development of multi scale physics, especially the factorization approach. The underlying idea, therefore, is the discussion on the nature of doubly charmed pentaquark in the paper.

Doubly charmed pentaquark $P(cc\bar qqq)$ can be dealt with the compact pentaquark interpretation. The diquark with two charmed quarks $cc$ forms a color triplet $\bar{3}_c$ spin-1 state, as suggested by perturbative arguments. While the light diquark $qq$ forms a color triplet $\bar{3}_c$ spin-0 state, as the hypothesis of ``good" diquark. At this stage, we employ the triquark-diquark model~\cite{Ali:2019npk,Ali:2019clg,Maiani:2014aja}, which heavy diquark $[cc]_{\bar 3}$ combines with antiquark $q_{\bar 3}$ to form a triquark system $[[cc]_{\bar3}q_{\bar 3}]_{3}$, and then, merging with light diquark $[qq']_{\bar 3}$ to form the doubly charmed pentquark state. Once applying the effective Hamiltonian of mass spectrum, the masses of pentaquark would be achieved. In addition, the lifetimes can be estimated with the implement of operator product expansion(OPE) approach~\cite{Lenz:2014jha,Ali:2018xfq}. Upon the heavy quark expanding and optical theorem, we can directly determine the lifetimes at the next-to-leading order of $P(cc\bar qqq)$ with the different parities.

The flavor SU(3) symmetry is a convincing tool to analyze production and decay behaviours of hadrons, which has been successfully applied to the meson or baryon system~\cite{Savage:1989ub,Gronau:1995hm,He:1998rq,Chiang:2004nm,Li:2007bh,Wang:2009azc,Cheng:2011qh,Hsiao:2015iiu,Lu:2016ogy,He:2016xvd,Wang:2017vnc,Wang:2017azm,Shi:2017dto,He:2018php,Shi:2020gfp,Li:2021rfj}. Though the SU(3) breaking effects in charm quark transition might  be sizable, the results can still describe the experimental data well in a global viewpoint. The doubly charmed pentaquark can be produced from triply charmed baryon $\Omega_{ccc}$ with one charmed quark weak decays, $c\to q\bar qq$. Moreover, the decays of pentaquark similarly give priority to the charmed quark decays. Therefore, we will force on the charmed quark weak decays both in the production and decays processes. To be more explicit, one can write down the Hamiltonian at the hadron level  with   hadron fields and transition operators. Some limited amount of input  parameters will be introduced  to describe the non-perturbative transitions. With the SU(3) amplitudes,  one can obtain   relations between  decay widths of different processes,  which can be examined  in experiment. Such an analysis is also helpful to identify the decay modes that will be mostly useful to discover the  doubly charmed pentaquark state $P(c c\bar qqq)$. Since the SU(3) analysis is based on the light quark symmetry, thus the analytical results can work well in all states with $P({cc}{\bar qqq})$ flavor constituents, for instance, the states with molecular or diquark picture.

The rest of this paper is organized as follows. In Sec.~\ref{sec:lifetime}, we discuss the mass spectrums and lifetimes of doubly charmed pentaquark. Section III is devoted to discuss the production and decays behaviours, which including two body production processes, mesonic semi- and non-leptonic decays. In section V, we present a collection of the golden channels. We make a short summary in the end.

\section{Mass and Lifetime of $P(c c\bar qqq)$ state}
\renewcommand\thesubsection{(\roman{subsection})}
\label{sec:lifetime}
\subsection{Mass}
\renewcommand\thesubsection{(\roman{subsection})}
\label{sec:mass}
We study the mass spectrum of S-wave pentaquark states $P_{cc\bar qqq}(q=u,d,s)$ in the framework of non-relativity doubly heavy triquark-diquark model~\cite{Ali:2019npk,Ali:2019clg,Maiani:2014aja}. Under the picture of diquark consisted of two light quarks $qq'$ in color $\bar 3$ state and triquark consisted of doubly heavy diquark $cc$ in color $\bar 3$ plus a light anti-quark $\bar q$ in color $\bar 3$, which labeled as $[[cc]_{c\bar3}[\bar q]_{c\bar 3}]_{c 3}[qq']_{c\bar 3}$, the effective Hamiltonian of mass spectrum can then be written as
\begin{eqnarray}
\label{eq:massh}
\mathcal{H}=M_0+H_{hd}+H_{ld},
\end{eqnarray}
with the constituent mass of diquark and triquark is $M_0$, in addition, the interaction of triquark $H_{hd}$ contains the coupling of spin-spin $S_c\cdot S_c$ in the heavy diquark $[cc]_{\bar 3}$, and the coupling with the triquark $S_c\cdot S_{\bar q}$. The last term $H_{ld}$ provides the interaction between two light quark $q$ and $q'$ in diquark, and all possible spin-spin interactions among the light diquark $[qq']_{\bar 3}$ and triquark $[[cc]_{\bar 3}{\bar q}_{\bar 3}]_{3}$. We respectively give the forms as follows.
\begin{eqnarray}
M_0&=&m_{cc}+m_{qq'}+m_{\bar q},\nonumber\\
H_{hd}&=&2 (\mathcal{K}_{cc})_{\bar 3} (S_c\cdot S_c)+4 (\mathcal{K}_{c\bar q})(S_c\cdot S_{\bar q}),\\\nonumber
H_{ld}&=&2(\mathcal{K}_{qq'})_{\bar3} (S_{q}\cdot S_{q'})+4(\mathcal{K}_{cq})_{\bar 3}(S_c\cdot S_q)+4(\mathcal{K}_{cq'})_{\bar 3}(S_c\cdot S_{q'})+2(\mathcal{K}_{q\bar q})(S_q\cdot S_{\bar q})+2(\mathcal{K}_{q'\bar q})(S_{q'}\cdot S_{\bar q}).
\end{eqnarray}

In the doubly heavy triquark-diquark system, the suggested spin of doubly heavy charmed diquark is $S_{cc}=1$. Similarily, the spin of ``good" light diquark in the S-wave pentaquark state $P_{cc\bar q qq}$ is chosen as $S_{qq'}=0$~\cite{Jaffe:2003sg}. Accordingly, we can write directly the possible configuration of S-wave pentaquark $P_{cc\bar qqq}$, signed as $|S_{cc},S_t,L_t; S_{qq'},L_{qq'};S,L\rangle$.
\begin{eqnarray}
&|1,\frac{1}{2},0;0,0;\frac{1}{2},0\rangle &\ \ J^P=\frac{1}{2}^-,\nonumber\\
&|1,\frac{3}{2},0;0,0;\frac{3}{2},0\rangle &\ \ J^P=\frac{3}{2}^-.
\end{eqnarray}
Here, the light diquark $qq'$ can be any one of the constituents $(ud,du,us,su,ds,sd)$. Except for the spin of doubly heavy diquark $S_{cc}$ and light diquark $S_{qq'}$, the orbital angular momentum $L_t=L_{qq'}=L=0$, the spin of triquark ($[cc] \bar q$) $S_t$ turn out to be $\frac{1}{2}$ or $\frac{3}{2}$.

The two states with parity $J^P=\frac{1}{2}^-$ and $J^P=\frac{3}{2}^-$, sandwiching the effective mass Hamiltonian Eq.~\ref{eq:massh}, then yield the mass spectrum matrix of S-wave pentaquark $P_{cc\bar qqq}$:
\begin{eqnarray}
\mathcal{M}^{S=0}_{J=1/2,3/2}&=&m_{cc}+m_{\bar q}+m_{qq'}+\frac{1}{2}(\mathcal{K}_{cc})_{\bar 3}-\frac{3}{2}(\mathcal{K}_{qq'})_{\bar3}\nonumber\\
&&+\left(\begin{array}{cc} -2\mathcal{K}_{c\bar q}+\frac{1}{4}(\mathcal{K}_{cq})_{\bar 3}+\frac{1}{4}(\mathcal{K}_{cq'})_{\bar 3}&\frac{1}{2}\mathcal{K}_{c\bar q}+\frac{1}{12}(1+\sqrt{10})(\mathcal{K}_{q\bar q}+\mathcal{K}_{q'\bar q})\\ \frac{1}{2}\mathcal{K}_{c\bar q}+\frac{1}{12}(1+\sqrt{10})(\mathcal{K}_{q\bar q}+\mathcal{K}_{q'\bar q})&\mathcal{K}_{c\bar q}   \end{array}\right).
\end{eqnarray}
In particular, the determination of spin-spin interaction between three spins inside the triquark, i.e., $S_c\cdot S_{\bar q}$, can be drawn by the Wigner 6j-symbols. Further more, for the interaction between triquark and light diquark, such as, $S_{c}\cdot S_{q}$ and $S_{q}\cdot S_{\bar q}$, it is convenient to utilize the Wigner 9j-symbols to describe the recouplings. The remaining step is the numerical operation. With regard to the work, we choose the values of spin-spin coupling given as~\cite{Ali:2009pi}: $(\mathcal{K}_{cc})_{\bar 3}=57$ MeV, $(\mathcal{K}_{qq'})_{\bar 3}=98$ MeV, $(\mathcal{K}_{su/d})_{\bar 3}=59$ MeV, $(\mathcal{K}_{cu/d})_{\bar 3}=15$ MeV, $(\mathcal{K}_{cs})_{\bar 3}=50$ MeV, $\mathcal{K}_{c\bar u/\bar d}=72$ MeV, $\mathcal{K}_{u\bar d/d\bar u}=318$ MeV, $\mathcal{K}_{s\bar d/\bar u}=200$ MeV. We adopt the mass of quark and diquark~\cite{Ali:2009pi}, for instance, $m_{u/d}=362$ MeV, $m_s=540$ MeV, $m_c=1.667$ GeV, $m_{cc}\sim2m_c$, $m_{ud}=576$ MeV, $m_{sq}=800$ MeV. Certainly, one should consider the uncertainty in these couplings and masses. We assign the couplings to be $10\%$ of each value, so as the mass of heavy charmed diquark.

We diagonalize the mass matrix and obtain the split mass of  pentaquark  $P_{cc\bar qqq}$ shown in Tab.~\ref{tab:mass}. As a contrast, we show the results from chiral effective theory(ChEFT), quark model with color-magnetic interaction(CMI) and QCD sum rule(QCDSR). Moreover, the lowest strong thresholds are placed at the end of the table. From which we may find that the center mass of pentaquark $P_{cc\bar s ud}$ with parity $\frac{1}{2}^-$ is higher than the lowest strong threshold $\Xi_{ccq}\bar{K}$ about 23 MeV. Therefore, it is possible a stable bound state against the strong interaction. The conclusion is consistent with the Refs~\cite{Guo:2017vcf,Park:2018oib}. Still, it is worth noting, the difference value much smaller than the uncertainties in our work, this may play an increasingly role at the final conclusion. In addition, the mass of P($\frac{1}{2}^-$) with the constituents $cc\bar u su, cc\bar u sd, cc\bar d s u$ and $cc\bar d s d$ are higher slightly than the lowest strong thresholds about $10$ MeV, which may lead the final options open for the considerable uncertainties from couplings $\mathcal{K}_{cc}$ and diquark mass $m_{cc}$. Nevertheless, the work have suggested the potential stable state $P_{cc\bar s ud}(\frac{1}{2}^-)$, which should be further confirmed in experimental sides. In the following, we will intimately discuss the production and decay modes of pentaquark $cc\bar qqq$ under the SU(3) symmetry analysis.
\begin{table}
  \centering
  \caption{The split mass of  pentaquark  $P_{cc\bar qqq}$ with the S-wave in the parity of $J^P=\frac{1}{2}^-,\frac{3}{2}^-$, in the unit of GeV. $n$ represent the light quark $u,d$. $I$ is the isospin of pentaquark.}\label{tab:mass}
 	\begin{tabular}{|l|c|c|c|c|c|}\hline\hline
	\multirow{2}*{Pentaquark($J^P,I$)} &  \multicolumn{4}{c|}{Mass} 	& \multirow{2}*{threshold}  \\ \cline{2-5}
	&  this work  & ChEFT \cite{Guo:2017vcf} &CMI \cite{Zhou:2018bkn} &QCDSR \cite{Wang:2018lhz} &       \\\hline
  $cc\bar s nn (\frac{1}{2}^-,0)$ &  $4.092\pm{0.298}$ &  3.957   & 4.702 & - &  $\Xi_{ccq}\bar{K}$ \\\hline
  $cc\bar n nn (\frac{1}{2}^-,\frac{1}{2})$ &  $3.841\pm{0.290}$ &  3.816  & 4.578 & 4.21 & $\Xi_{ccq}\pi$ \\\hline
  $cc\bar n sn (\frac{1}{2}^-,1/0)$ &  $4.125\pm{0.301}$ &  4.112  & 4.854  &- & $\Xi_{ccq}K$ \\\hline
  $cc\bar s sn (\frac{1}{2}^-,\frac{1}{2})$ &  $4.409\pm{0.307}$ & 3.816 & 4.968 &-   & $\Omega_{ccq}K$ \\\hline
  \hline
  $cc\bar s nn (\frac{3}{2}^-,0)$ &  $4.496\pm{0.338}$ &   - & 4.355 & - & $\Xi_{ccq}^*\bar{K}$\\\hline
  $cc\bar n nn (\frac{3}{2}^-,\frac{1}{2})$ &  $4.393\pm{0.340}$ &-    & 3.970   &  4.27 & $\Xi_{ccq}^*\pi$  \\\hline
  $cc\bar n sn (\frac{3}{2}^-,1/0)$ &  $4.600\pm{0.348}$ &  -  &  4.802  &  -  &  $\Xi_{ccq}^*K$  \\\hline
  $cc\bar s sn (\frac{1}{2}^-,\frac{1}{2})$ &  $4.762\pm{0.342}$ & -  &  4.955 &-  & $\Omega_{ccq}^*K$ \\\hline
  \end{tabular}
\end{table}
\subsection{Lifetimes}
\renewcommand\thesubsection{(\roman{subsection})}
\label{sec:lifetimes}
In the part, we roughly study the lifetimes of pentaquark $P_{cc\bar qqq}$ with the parity $J^P=\frac{1}{2}^-,\frac{3}{2}^-$, under the framework of operator product expansion(OPE) technique. As always, the decays widths can be expressed as
\begin{eqnarray}
&\Gamma(P_{cc\bar qqq}\to X)=\frac{1}{2m_P} \sum_X \int \prod_i \Big[ \frac{d^3 \overrightarrow{p}_i}{(2\pi)^3 2E_i}\Big] (2\pi)^4 \delta^4 p_P-\sum_i p_i) N\sum_{\lambda}|\langle X|\mathcal{H}| P_{cc\bar qqq}\rangle|^2,
\end{eqnarray}
therein, $m_P$, $\lambda$ and $p_P^{\mu}$ are the mass, spin and four-momentum  of pentaquark $P_{cc\bar qqq}$ respectively. $\mathcal{H}$ can match with that of electro-weak effective Hamiltonian $\mathcal{H}_{eff}^{ew}$. The coefficient $N=(\frac{1}{2},\frac{1}{4})$, accordingly corresponds with spin parity $J^P=(\frac{1}{2}^-, \frac{3}{2}^-)$ of pentaquark. Based on the optical theorem and the heavy quark expanding(HQE), the decay widths from leading dimension contribution can be deduced as
\begin{eqnarray}
&\Gamma(P_{cc\bar qqq} \to X)= \frac{G_F^2m_c^5}{192\pi^3} |V_{CKM}|^2 N \sum_{\lambda} c_{3,c} \frac{\langle P_{cc\bar qqq}|\bar c c|P_{cc\bar qqq}\rangle}{2m_P},
\end{eqnarray}
where $G_F$ is the Fermi constant, $V_{CKM}$ is the CKM element, the coefficients $c_{3,c}$ is the perturbative coefficient of HQE. Further more, the heavy quark matrix element is corresponding with charm number of pentaquark state.
\begin{eqnarray}
\sum_{\lambda} \frac{\langle P_{cc\bar qqq}|\bar c c|P_{cc\bar qqq}\rangle}{2m_P}=2+\mathcal{O}(1/m_c).
\end{eqnarray}
Consequently, we reach the decay widths and lifetimes under the leading and next-to-leading order,
\begin{eqnarray}
&\Gamma(P(\frac{1}{2}^-))=\left\{\begin{array}{l} (7.68^{+0.88}_{-0.88})\times 10^{-13}  {\rm GeV} ,
\;\text{LO} \\ (1.42^{+0.19}_{-0.19})\times 10^{-12} \ {\rm GeV}  ,\; \text{NLO} \end{array}\right. ,
&\Gamma(P(\frac{3}{2}^-))=\left\{\begin{array}{l} (3.84^{+0.44}_{-0.44})\times 10^{-13}  {\rm GeV} ,
\;\text{LO} \\ (7.08^{+0.95}_{-0.95})\times 10^{-13} \ {\rm GeV}  ,\; \text{NLO} \end{array}\right. ,\nonumber\\
&\tau(P(\frac{1}{2}^-))=\left\{\begin{array}{l} (8.58^{+1.11}_{- 0.88})\times 10^{-13}  s ,
\;\text{LO} \\ (4.65^{+0.71}_{-0.55})\times 10^{-13
}
 s  ,\; \text{NLO} \end{array}\right. ,
&\tau(P(\frac{3}{2}^-))=\left\{\begin{array}{l} (1.72^{+0.22}_{-0.18})\times 10^{-12}  s ,
\;\text{LO} \\ (0.93^{+0.14}_{-0.11})\times 10^{-12}  s  ,\; \text{NLO} \end{array}\right. .
\end{eqnarray}
In this work, the heavy quark mass $m_c=1.4\ {\rm GeV}$, the perturbative short-distance coefficient have been determined as $c_{3,c}=6.29\pm0.72$ at the leading order(LO), and $c_{3,c}=11.61\pm1.55$ at the next-to-leading order(NLO)~\cite{Lenz:2014jha}.

\section{SU(3) analysis}
\renewcommand\thesubsection{(\Roman{subsection})}
\label{sec:weak_decay}
We will discuss the possible production and decay modes of pentaquark with the quark constituent of $cc\bar qqq$ in this section. The production can be achieved by the study about weak decays of triply heavy baryon $\Omega_{ccc}$. At the stage of decay modes, we focus on the explores of stable pentaquark candidates, which give priority to the weak decays similarly.

The weak interaction of production and decays for the pentaquark states, with the transition $c\to q\bar qq$, can be classified  by the quantities of CKM matrix elements.
\begin{itemize}
\item For the case of $c$ quark semi-leptonic decays,
\begin{eqnarray}
 c\to   d/ s  \ell^+   \nu_{\ell}.
\end{eqnarray}
The general electro-weak Hamiltonian can be expressed as
\begin{eqnarray}
 {\cal H}_{eff} &=& \frac{G_F}{\sqrt2} \left[V_{cq}^* \bar q  \gamma^\mu(1-\gamma_5)c \bar \nu_{\ell} \gamma_\mu(1-\gamma_5) \ell\right] +h.c.,
\end{eqnarray}
with $q=(d,s)$, in which the transition operator of $c\to q \ell^+ \nu_{\ell}$ forms a SU(3) triplet $H_{  3}$, and $(H_{  3})_1=0,~(H_{  3})_2=V_{cd}^*,~(H_{  3})_3=V_{cs}^*$.
\item  For the case of $c$ quark non-leptonic decays, we classify the transitions into three groups.
\begin{eqnarray}
c\to  s \bar d  u, \;
c\to  u \bar d  d/\bar s s, \;
c\to  d \bar s  u, \;
\end{eqnarray}
which are Cabibbo allowed, singly Cabibbo suppressed, and doubly Cabibbo suppressed transitions respectively.
The transition $c \to  q_1  \bar q_2  q_3$  can be decomposed as ${\bf  3}\otimes {\bf 3}\otimes {\bf
\bar3}={\bf  3}\oplus {\bf  3}\oplus {\bf \bar 6}\oplus {\bf 15}$.
Here, we offer the nonzero SU(3) tensor components of Cabibbo allowed transition given as
\begin{eqnarray}
(H_{ 6})_{31}^2=-(H_{6})_{13}^2=1,\;\;\;
 (H_{\overline {15}})_{31}^2= (H_{\overline {15}})_{13}^2=1.\label{eq:H3615_c_allowed}
\end{eqnarray}
As the transition of $\bar c\to \bar u d\bar d$ and $\bar c\to \bar u  s\bar s$ with singly Cabibbo suppressed, the combination of tensor components are corresponding with the overall CKM factor, which defined as $V_{cs}^* V_{us}=sin(\theta_C)$.
\begin{eqnarray}
(H_{6})_{31}^3 =-(H_{6})_{13}^3 =(H_{ 6})_{12}^2 =-(H_{ 6})_{21}^2 =\sin(\theta_C),\nonumber\\
 (H_{\overline {15}})_{31}^3= (H_{\overline {15}})_{13}^3=-(H_{\overline {15}})_{12}^2=-(H_{\overline {15}})_{21}^2= \sin(\theta_C).\label{eq:H3615_cc_singly_suppressed}
\end{eqnarray}
For the doubly Cabibbo suppressed transition  $\bar c\to \bar d  s \bar u$, we have
\begin{eqnarray}
(H_{ 6})_{21}^3=-(H_{ 6})_{12}^3=-\sin^2\theta_C,\;\;
 (H_{\overline {15}})_{21}^3= (H_{\overline {15}})_{12}^3=-\sin^2\theta_C. \label{eq:H3615_c_doubly_suprressed}
\end{eqnarray}
\end{itemize}

\subsection{Particle Multiplets in SU(3)}
\label{sec:particle_multiplet}
The pentaquark with the quark constituents ${c c}{\bar qqq}$ contain three light quark and thus can form an SU(3) $\mathbf{\bar 6}$ and an SU(3) $\mathbf{15}$, labeled as $P_6$ and $P_{15}$ respectively.
We give the SU(3) representations $P_6$ as follows.
\begin{eqnarray}
&&(P_{6})^{[2,3]}_1= \frac{1}{\sqrt{2}} P_{\bar uds}^0,(P_{6})^{[3,1]}_2= \frac{1}{\sqrt{2}} P_{\bar dsu}^{++},(P_{6})^{[1,2]}_3= \frac{1}{\sqrt{2}} P_{\bar sud}^{++}\\\nonumber
&&(P_{6})^{[1,2]}_1=(P_{6})^{[2,3]}_3= \frac{1}{2} P_{(\bar uu,\bar{s}s)d}^+,(P_{6})^{[3,1]}_1=(P_{6})^{[2,3]}_2= \frac{1}{2} P_{(\bar{u}u,\bar{d}d)s}^{+},\\\nonumber
&& (P_{6})^{[1,2]}_2=(P_{6})^{[3,1]}_3= \frac{1}{2} P_{(\bar{d}d,\bar{s}s)u}^{++}.
\end{eqnarray}
It should be noted that the tensor representation, for instance, $(P_{6})^{[2,3]}_1$ with two antisymmetry upper indices $[2,3]$, can be reformulated by the SU(3) invariant tensor $\varepsilon^{ijk}$ written as $(P_{6})^{[2,3]}_1=(P_6)_{\{1,1\}} \varepsilon^{123}$. The component fields in $15$ states are
\begin{eqnarray}
&&(P_{15})^{\{2,3\}}_1= \frac{1}{\sqrt{2}} {P'}_{ds\bar u}^0,(P_{15})^{\{3,1\}}_2= \frac{1}{\sqrt{2}} {P'}_{su\bar d}^{++},(P_{15})^{\{1,2\}}_3= \frac{1}{\sqrt{2}} P_{ud\bar s}^{++}, \\\nonumber
&&(P_{15})^{\{1,1\}}_1=\frac{{P'}_{\pi u}^{++}}{\sqrt2}+\frac{{P'}_{\eta u}^{++}}{\sqrt6}, (P_{15})^{\{1,2\}}_1=\frac{1}{\sqrt{2}}(\frac{{P'}_{\pi d}^{+}}{\sqrt{2}}+\frac{{P'}_{\eta d}^{+}}{\sqrt6}), (P_{15})^{\{1,3\}}_1=\frac{1}{\sqrt2}(\frac{{P'}^{+}_{\pi s}}{\sqrt2}+\frac{{P'}^{+}_{\eta s}}{\sqrt6}),\\\nonumber
&& (P_{15})^{\{2,1\}}_2=\frac{1}{\sqrt2}(-\frac{{P'}^{++}_{\pi u}}{\sqrt2}+\frac{{P'}^{++}_{\eta u}}{\sqrt6}), (P_{15})^{\{2,2\}}_2=(-\frac{{P'}^{+}_{\pi d}}{\sqrt2}+\frac{{P'}^{+}_{\eta d}}{\sqrt6}), (P_{15})^{\{2,3\}}_2=\frac{1}{\sqrt2}(-\frac{{P'}^{+}_{\pi s}}{\sqrt2}+\frac{{P'}^{+}_{\eta s}}{\sqrt6}), \\\nonumber
&& (P_{15})^{\{3,1\}}_3=-\frac{1}{\sqrt3}{P'}^{++}_{\eta u},  (P_{15})^{\{3,2\}}_3=-\frac{1}{\sqrt3}{P'}^{+}_{\eta d}, (P_{15})^{\{3,3\}}_3=-\frac{1}{\sqrt3}{P'}^{+}_{\eta s} \\\nonumber
&& (P_{15})^{\{2,2\}}_1={P'}^{0}_{dd\bar u}, (P_{15})^{\{3,3\}}_1={P'}^{0}_{ss\bar u}, (P_{15})^{\{1,1\}}_2={P'}^{+++}_{uu\bar d}, \\\nonumber
&& (P_{15})^{\{3,3\}}_2={P'}^{+}_{ss\bar d}, (P_{15})^{\{1,1\}}_3={P'}^{+++}_{uu\bar s}, (P_{15})^{\{2,2\}}_3={P'}^{+}_{dd\bar s}.
\end{eqnarray}
In addition, the baryons with the components $qqq$, can form an SU(3) octet $T_8$ and an SU(3) decuplet $T_{10}$. The octet has the expression
\begin{eqnarray}
T_8= \left(\begin{array}{ccc} \frac{1}{\sqrt{2}} \Sigma^0+\frac{1}{\sqrt{6}} \Lambda^0 &  \Sigma^+  &  p   \\  \Sigma^- &  -\frac{1}{\sqrt{2}} \Sigma^0+\frac{1}{\sqrt{6}} \Lambda^0&  n \\  \Xi^-   &  \Xi^0 & -\sqrt{\frac{2}{3}} \Lambda^0
  \end{array} \right),
\end{eqnarray}
and the light decuplet is given as
\begin{eqnarray}
(T_{10})^{111} &=& \Delta^{++},\;\;\; (T_{10})^{112}= (T_{10})^{121}=(T_{10})^{211}= \frac{1}{\sqrt3} \Delta^{+},\nonumber\\
(T_{10})^{222} &=& \Delta^{-},\;\;\; (T_{10})^{122}= (T_{10})^{212}=(T_{10})^{221}= \frac{1}{\sqrt3}  \Delta^0, \nonumber\\
(T_{10})^{113} &=& (T_{10})^{131}=(T_{10})^{311}= \frac{1}{\sqrt3}  \Sigma^{\prime+},\;\;(T_{10})^{223} = (T_{10})^{232}=(T_{10})^{322}= \frac{1}{\sqrt3} \Sigma^{\prime-},\nonumber\\
(T_{10})^{123} &=& (T_{10})^{132}=(T_{10})^{213}=(T_{10})^{231}=(T_{10})^{312}=(T_{10})^{321}= \frac{1}{\sqrt6} \Sigma^{\prime0},\nonumber\\
(T_{10})^{133} &=& (T_{10})^{313}=(T_{10})^{331}= \frac{1}{\sqrt3} \Xi^{\prime0},\;\;(T_{10})^{233} = (T_{10})^{323}=(T_{10})^{332}= \frac{1}{\sqrt3}  \Xi^{\prime-}, \nonumber\\
(T_{10})^{333}&=&  \Omega^-.
\end{eqnarray}
Consistently, the singly charmed baryons $cqq$ are expected to form a anti-triplet and a sextet, respectively as~\cite{Xing:2018bqt,Xing:2019wil}
\begin{eqnarray}
 T_{\bf{c\bar 3}}= \left(\begin{array}{ccc} 0 & \Lambda_c^+  &  \Xi_c^+  \\ -\Lambda_c^+ & 0 & \Xi_c^0 \\ -\Xi_c^+   &  -\Xi_c^0  & 0
  \end{array} \right), \;\;
 T_{\bf{c
 6}} = \left(\begin{array}{ccc} \Sigma_c^{++} &  \frac{1}{\sqrt{2}}\Sigma_c^+   & \frac{1}{\sqrt{2}} \Xi_c^{\prime+}\\
  \frac{1}{\sqrt{2}}\Sigma_c^+& \Sigma_c^{0} & \frac{1}{\sqrt{2}} \Xi_c^{\prime0} \\
  \frac{1}{\sqrt{2}} \Xi_c^{\prime+}   &  \frac{1}{\sqrt{2}} \Xi_c^{\prime0}  & \Omega_c^0
  \end{array} \right)\,.
\end{eqnarray}

\begin{figure}
  \centering
  \includegraphics[width=0.80\columnwidth]{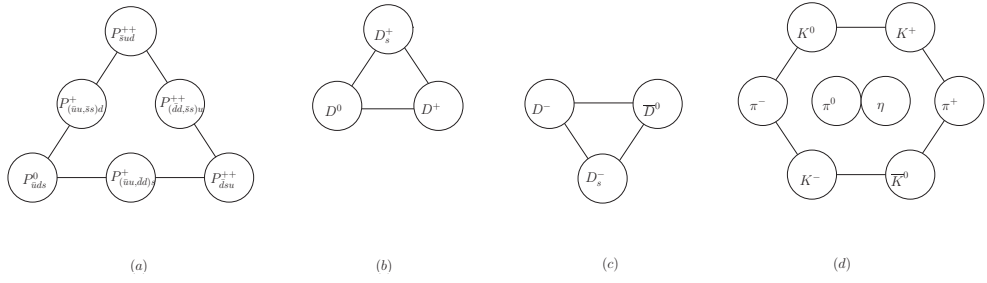}\\
  \caption{The weight diagram for the pentaquark sextet $P_6$ is given in (a). Besides, The weight diagrams of mesons, provided in (b,c,d), charmed mesons anti-triplet and triplet to be (b,c), light meson octet to be (d).}\label{fig:multiplet-pentaquark}
\end{figure}

\begin{figure}
  \centering
  \includegraphics[width=1.0\columnwidth]{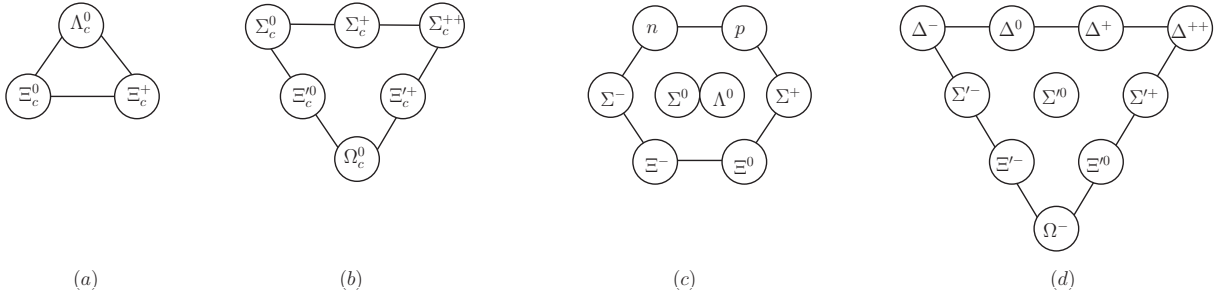}\\
  \caption{The weight diagrams for the singly charmed baryons are given in (a,b), which triplet $T_{c\bar3}$ to be (a), sextet $T_{c6}$ to be (b). The light baryon multiplets are $T_8, T_{10} $ shown in (c,d) respectively.}\label{fig:multiplet-baryons}
\end{figure}

In the meson sector, singly charmed mesons form an SU(3) triplet or anti-triplet, light mesons form an octet plus singlet, all multiplets are collected as
\begin{eqnarray}
 M_{8}=\begin{pmatrix}
 \frac{\pi^0}{\sqrt{2}}+\frac{\eta}{\sqrt{6}}
 &\pi^+ & K^+\\
 \pi^-&-\frac{\pi^0}{\sqrt{2}}+\frac{\eta}{\sqrt{6}}&{K^0}\\
 K^-&\bar K^0 &-2\frac{\eta}{\sqrt{6}}
 \end{pmatrix},
D_i^T = \left(\begin{array}{c}  D^0  \\  D^+  \\  D^+_s
\end{array}\right)\,,\;\;
\overline D^i= \left(\begin{array}{c} \overline D^0  \\  D^-  \\  D^-_s
\end{array}\right).
\end{eqnarray}
For completeness, we also draw the weight diagrams of multiplets, shown in Fig.~\ref{fig:multiplet-pentaquark} and  Fig.~\ref{fig:multiplet-baryons}.

\subsection{Production of  $P_{c c\bar qqq}$ from $\Omega_{ccc}$}
\begin{figure}
  \centering
  \includegraphics[width=0.55\columnwidth]{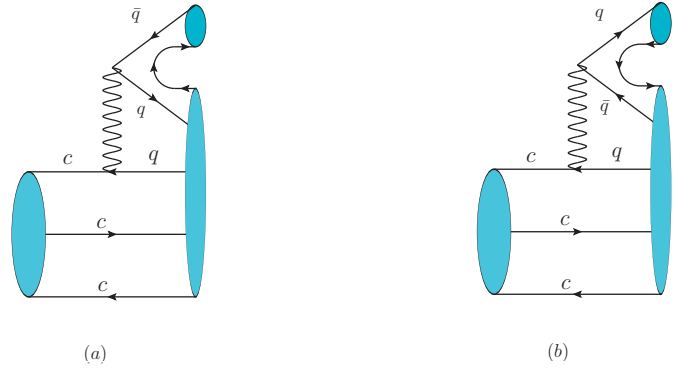}\\
  \caption{Two typical Feynman diagrams for the production of doubly heavy pentaquark $P_{cc\bar qqq}$ from the triply charmed baryon $\Omega_{ccc}$. The production depends on the transition $c\to q\bar qq$ in quark level, which leads to similar topologies (a,b) including one light meson and doubly heavy pentaquark in final states.}\label{fig:production}
\end{figure}

\renewcommand\thesubsection{(\roman{subsection})}
\label{sec:production}
The pentaquark $P_{cc\bar qqq}$ can be produced by the weak decays of triply charmed baryon $\Omega_{ccc}$, once one charmed quark decays $c\to q\bar qq$ in baryon. The typical Feynman diagrams with two final states are shown in Fig.~\ref{fig:production}. In particular, the final states include the pentaquark we wanted and a light meson.
Within the framework of SU(3) symmetry analysis, we then construct the Hamiltonian at the hadronic level, which straightly written as
\begin{eqnarray*}
\mathcal{H}&=&a_1 \Omega_{ccc} (H_6)^{[ij]}_{l} (P_6)_{[ij]}^k (M)_k^l+a_2 \Omega_{ccc} (H_6)^{[ij]}_{k} (P_6)_{[il]}^k (M)_j^l+a_3 \Omega_{ccc} (H_{15})^{[ij]}_{k} (P_6)_{[il]}^k (M)_j^l,\\
\mathcal{H}&=&b_1 \Omega_{ccc} (H_{6})^{[ij]}_{k} (P_{15})_{\{il\}}^k (M)_j^l+b_2 \Omega_{ccc} (H_{15})^{\{ij\}}_{l} (P_{15})_{\{ij\}}^k (M)_k^l+b_3 \Omega_{ccc} (H_{15})^{[ij]}_{k} (P_{15})_{\{il\}}^k (M)_j^l.
\end{eqnarray*}
where the coefficients, such as $a_1,a_2,b_1,b_2,...$, represent the non-perturbative parameters. The SU(3) representations of corresponding hadrons and transitions above, permit us to expand the Hamiltonian into diverse decay channels, revealed with combinations of non-perturbative coefficients. We forward and collect the possible production channels of $P_6(cc\bar{q}qq)$ states in Tab.~\ref{tab:Omegaccc_P6}. Especially, the Cabibbo allowed, singly Cabibbo suppressed and doubly Cabibbo suppressed are displayed respectively. In addition, the channels corresponding with $P_{15}(cc\bar{q}qq)$ states are placed into Tab.~\ref{tab:Omegaccc_P15}. Referring to the amplitudes of production channels above, we reduce the relations between different channels.
\begin{eqnarray*}
&&\Gamma(\Omega_{ccc}^{++}\to P_{\bar{d}su}^{++} \pi^0 )= \Gamma(\Omega_{ccc}^{++}\to P_{\{\bar{u}u,\bar{d}d\}\bar{s}}^{+} \pi^+ ),
 \Gamma(\Omega_{ccc}^{++}\to P_{\bar{d}su}^{++} K^0 )= \Gamma(\Omega_{ccc}^{++}\to P_{\bar{s}ud}^{++} \overline K^0 ),\\
&&\Gamma(\Omega_{ccc}^{++}\to P_{\{\bar{u}u,\bar{s}s\}\bar{d}}^{+} \pi^+ )= \Gamma(\Omega_{ccc}^{++}\to P_{\{\bar{u}u,\bar{d}d\}\bar{s}}^{+} K^+ ),
 \Gamma(\Omega_{ccc}^{++}\to P_{\{\bar{d}d,\bar{s}s\}\bar{u}}^{++} \pi^0 )= 3\Gamma(\Omega_{ccc}^{++}\to P_{\{\bar{d}d,\bar{s}s\}\bar{u}}^{++} \eta_q ),\\
&&\Gamma(\Omega_{ccc}^{++}\to {P'}_{ud\bar{s}}^{++} \overline K^0 )= \Gamma(\Omega_{ccc}^{++}\to {P'}_{su\bar{d}}^{++} K^0 ),
 \Gamma(\Omega_{ccc}^{++}\to {P'}_{su\bar{d}}^{++} \pi^0 )= \Gamma(\Omega_{ccc}^{++}\to {P'}_{\pi s}^{+} \pi^+ ),\\
&&\Gamma(\Omega_{ccc}^{++}\to {P'}_{\pi u}^{++} \pi^0 )= \Gamma(\Omega_{ccc}^{++}\to {P'}_{\eta u}^{++} \eta_q ),
 \Gamma(\Omega_{ccc}^{++}\to {P'}_{\pi u}^{++} K^0 )= \Gamma(\Omega_{ccc}^{++}\to {P'}_{\pi d}^{+} K^+ ).
\end{eqnarray*}
The Cabibbo allowed production channel of $P_6(cc\bar{q}qq)$ and $P_{15}(cc\bar{q}qq)$ states $\Omega_{ccc}^{++}\to   P_{\{\bar{u}u,\bar{d}d\}\bar{s}}^{+}  \pi^+$, $\Omega_{ccc}^{++}\to   {P'}_{\eta s}^{+}  \pi^+  $, $\Omega_{ccc}^{++}\to   {P'}_{\pi s}^{+}  \pi^+  $ and $\Omega_{ccc}^{++}\to   {P'}_{ss\bar{d}}^{+}  K^+  $, receive the largest contribution, meanwhile, the charged final light mesons possess high detection efficiency. Consequently, we suggest that these channels can be the preference choices for the studying in the future experiment.
\begin{table}
\caption{The production of pentaquark $P_{6}(cc\bar{q}qq)$ from triply heavy baryon $\Omega_{ccc}$, sC represents the overall CKM factor $\sin(\theta_ C)$.}\label{tab:Omegaccc_P6}\begin{tabular}{|c|c|c|c|c|c|c|c}\hline\hline
channel & amplitude&channel &amplitude \\\hline
$\Omega_{ccc}^{++}\to   P_{\bar{d}su}^{++}  \pi^0  $ & $ \frac{1}{2} \left(-2 a_1+a_2+a_3\right)$&
$\Omega_{ccc}^{++}\to   P_{\bar{d}su}^{++}  \eta_q  $ & $ \frac{2 a_1-a_2+3 a_3}{2 \sqrt{3}}$\\\hline
$\Omega_{ccc}^{++}\to   P_{\{\bar{u}u,\bar{d}d\}\bar{s}}^{+}  \pi^+  $ & $ a_1-\frac{a_2}{2}-\frac{a_3}{2}$&
$\Omega_{ccc}^{++}\to   P_{\{\bar{d}d,\bar{s}s\}\bar{u}}^{++}  \overline K^0  $ & $ \frac{1}{2} \left(2 a_1-a_2+a_3\right)$\\\hline
\hline$\Omega_{ccc}^{++}\to   P_{\bar{d}su}^{++}  K^0  $ & $ \frac{\left(2 a_1-a_2+a_3\right) \text{sC}}{\sqrt{2}}$&
$\Omega_{ccc}^{++}\to   P_{\bar{s}ud}^{++}  \overline K^0  $ & $ \frac{\left(2 a_1-a_2+a_3\right) \text{sC}}{\sqrt{2}}$\\\hline
$\Omega_{ccc}^{++}\to   P_{\{\bar{u}u,\bar{s}s\}\bar{d}}^{+}  \pi^+  $ & $ \frac{ \left(2 a_1-a_2-a_3\right) \text{sC}}{2}$&
$\Omega_{ccc}^{++}\to   P_{\{\bar{u}u,\bar{d}d\}\bar{s}}^{+}  K^+  $ & $ \frac{\left(2 a_1-a_2-a_3\right) \text{sC}}{2}$\\\hline
$\Omega_{ccc}^{++}\to   P_{\{\bar{d}d,\bar{s}s\}\bar{u}}^{++}  \pi^0  $ & $ \frac{\left(-2 a_1+a_2+3 a_3\right) \text{sC}}{2 \sqrt{2}}$&
$\Omega_{ccc}^{++}\to   P_{\{\bar{d}d,\bar{s}s\}\bar{u}}^{++}  \eta_q  $ & $ \frac{\left(-2 a_1+a_2+3 a_3\right) \text{sC}}{2 \sqrt{6}}$\\\hline
\hline$\Omega_{ccc}^{++}\to   P_{\bar{s}ud}^{++}  \pi^0  $ & $ a_3 \left(-\text{sC}^2\right)$&
$\Omega_{ccc}^{++}\to   P_{\bar{s}ud}^{++}  \eta_q  $ & $ \frac{\left(2 a_1-a_2\right) \text{sC}^2}{\sqrt{3}}$\\\hline
$\Omega_{ccc}^{++}\to   P_{\{\bar{u}u,\bar{s}s\}\bar{d}}^{+}  K^+  $ & $ \frac{ \left(-2 a_1+a_2+a_3\right) \text{sC}^2}{2}$&
$\Omega_{ccc}^{++}\to   P_{\{\bar{d}d,\bar{s}s\}\bar{u}}^{++}  K^0  $ & $ -\frac{ \left(2 a_1-a_2+a_3\right) \text{sC}^2}{2}$\\\hline
\hline
\end{tabular}
\end{table}
\begin{table}
\caption{The production of pentaquark $P_{15}(cc\bar{q}qq)$ from triply heavy baryon $\Omega_{ccc}$.}\label{tab:Omegaccc_P15}\begin{tabular}{|c|c|c|c|c|c|c|c}\hline\hline
channel & amplitude \\\hline
$\Omega_{ccc}^{++}\to   {P'}_{su\bar{d}}^{++}  \pi^0  $ & $ \frac{1}{2} \left(b_1-2 b_2+b_3\right)$&
$\Omega_{ccc}^{++}\to   {P'}_{su\bar{d}}^{++}  \eta_q  $ & $ \frac{3 b_1+2 b_2-b_3}{2 \sqrt{3}}$\\\hline
$\Omega_{ccc}^{++}\to   {P'}_{\pi u}^{++}  \overline K^0  $ & $ \frac{1}{2} \left(b_1-b_3\right)$&
$\Omega_{ccc}^{++}\to   {P'}_{\eta u}^{++}  \overline K^0  $ & $ -\frac{b_1+4 b_2-b_3}{2 \sqrt{3}}$\\\hline
$\Omega_{ccc}^{++}\to   {P'}_{\eta s}^{+}  \pi^+  $ & $ \frac{b_1+2 b_2+b_3}{2 \sqrt{3}}$&
$\Omega_{ccc}^{++}\to   {P'}_{\pi s}^{+}  \pi^+  $ & $ -\frac{b_1}{2}+b_2-\frac{b_3}{2}$\\\hline
$\Omega_{ccc}^{++}\to   {P'}_{ss\bar{d}}^{+}  K^+  $ & $ b_1+b_3$&&\\\hline
\hline$\Omega_{ccc}^{++}\to   {P'}_{ud\bar{s}}^{++}  \overline K^0  $ & $ -\frac{\left(b_1+2 b_2-b_3\right) \text{sC}}{\sqrt{2}}$&
$\Omega_{ccc}^{++}\to   {P'}_{su\bar{d}}^{++}  K^0  $ & $ \frac{\left(b_1+2 b_2-b_3\right) \text{sC}}{\sqrt{2}}$\\\hline
$\Omega_{ccc}^{++}\to   {P'}_{\pi u}^{++}  \pi^0  $ & $ \frac{\left(b_1-b_2\right) \text{sC}}{\sqrt{2}}$&
$\Omega_{ccc}^{++}\to   {P'}_{\pi u}^{++}  \eta_q  $ & $ \frac{\left(b_2+b_3\right) \text{sC}}{\sqrt{6}}$\\\hline
$\Omega_{ccc}^{++}\to   {P'}_{\eta u}^{++}  \pi^0  $ & $ -\frac{\left(2 b_1-b_2+b_3\right) \text{sC}}{\sqrt{6}}$&
$\Omega_{ccc}^{++}\to   {P'}_{\eta u}^{++}  \eta_q  $ & $ \frac{\left(b_2-b_1\right) \text{sC}}{\sqrt{2}}$\\\hline
$\Omega_{ccc}^{++}\to   {P'}_{\eta d}^{+}  \pi^+  $ & $ -\frac{\left(3 b_1+2 b_2+3 b_3\right) \text{sC}}{2 \sqrt{3}}$&
$\Omega_{ccc}^{++}\to   {P'}_{\pi d}^{+}  \pi^+  $ & $ \frac{\left(b_1-2 b_2+b_3\right) \text{sC}}{2}$\\\hline
$\Omega_{ccc}^{++}\to   {P'}_{\eta s}^{+}  K^+  $ & $ -\frac{\sqrt{3} \left(3 b_1-2 b_2+3  b_3\right) \text{sC}}{6}$&
$\Omega_{ccc}^{++}\to   {P'}_{\pi s}^{+}  K^+  $ & $ \frac{\left(b_1+2 b_2+b_3\right) \text{sC}}{2}$\\\hline
$\Omega_{ccc}^{++}\to   {P'}_{uu\bar{s}}^{+++}  K^-  $ & $ \left(b_3-b_1\right) \text{sC}$&&\\\hline
\hline$\Omega_{ccc}^{++}\to   {P'}_{ud\bar{s}}^{++}  \pi^0  $ & $ b_1 \text{sC}^2$&
$\Omega_{ccc}^{++}\to   {P'}_{ud\bar{s}}^{++}  \eta_q  $ & $ \frac{\left(b_3-2 b_2\right) \text{sC}^2}{\sqrt{3}}$\\\hline
$\Omega_{ccc}^{++}\to   {P'}_{\pi u}^{++}  K^0  $ & $ b_2 \left(-\text{sC}^2\right)$&
$\Omega_{ccc}^{++}\to   {P'}_{\eta u}^{++}  K^0  $ & $ \frac{\left(b_1+b_2-b_3\right) \text{sC}^2}{\sqrt{3}}$\\\hline
$\Omega_{ccc}^{++}\to   {P'}_{\eta d}^{+}  K^+  $ & $ -\frac{\left(b_1-b_2+b_3\right) \text{sC}^2}{\sqrt{3}}$&
$\Omega_{ccc}^{++}\to   {P'}_{\pi d}^{+}  K^+  $ & $ b_2 \text{sC}^2$\\\hline
$\Omega_{ccc}^{++}\to   {P'}_{uu\bar{s}}^{+++}  \pi^-  $ & $ \left(b_1-b_3\right) \left(-\text{sC}^2\right)$&
$\Omega_{ccc}^{++}\to   {P'}_{dd\bar{s}}^{+}  \pi^+  $ & $ \left(b_1+b_3\right) \text{sC}^2$\\\hline
\hline
\end{tabular}
\end{table}
\subsection{Decay modes of pentaquark $P_6(cc\bar qqq)$}
\renewcommand\thesubsection{(\roman{subsection})}
\label{sec:nonleptonic_decay}
In this part, we will study the possible decays of pentaquark $P_{c c\bar qqq}$ states. Generally, the excited states $P_{15}$ can primarily decays into $P_6$ states. In that case, we need only to discuss the decay modes of $P_6$ states individually in the paper.
As first step, we consider the semileptonic decays of $P_6(cc\bar qqq)$ with the transition of $c \to d/s \ell^+ \nu$. Following the SU(3) analysis, the construction of the corresponding Hamiltonian is forward, which can be constructed as
\begin{eqnarray}
\label{eq:Hamiltonian1}
\mathcal{H}&=&a_1' (P_6)^{[ij]}_{k} (H_3)^k (\overline{T}_{c\bar3})_{[ij]} \ell \bar{\nu}.
\end{eqnarray}
The Hamiltonian including sextet pentaquark $P_6$ with two antisymmetry upper indices in initial state, and triplet singly charmed baryon $T_{c\bar3}$ with two antisymmetry lower indices in final states, can then be represented into a compact form: $a_1 (P_6)_{\{ij\}} (H_3)^i (\overline{T}_{c\bar3})^j \ell \bar{\nu}$. We expand the new Hamiltonian and obtain six channels, $\mathcal{M}(P_{\bar{d}su}^{++}\to \Xi_c^+ l^+\nu)= -\sqrt{2} \mathcal{M}(P_{\{\bar{u}u,\bar{d}d\}\bar{s}}^{+}\to \Xi_c^0 l^+\nu )=-\sqrt2\mathcal{M}(P_{\{\bar{d}d,\bar{s}s\}\bar{u}}^{++}\to \Lambda_c^+ l^+\nu)=-\frac{a_1 \left(V_{\text{cd}}\right){}^*}{\sqrt{2}}$, and $\mathcal{M}(P_{\bar{s}ud}^{++}\to \Lambda_c^+ l^+\nu)=\sqrt2 \mathcal{M}(P_{\{\bar{u}u,\bar{s}s\}\bar{d}}^{+}\to \Xi_c^0 l^+\nu)=-\sqrt2\mathcal{M}(_{\{\bar{d}d,\bar{s}s\}\bar{u}}^{++}\to \Xi_c^+ l^+\nu)=\frac{a_1 \left(V_{\text{cs}}\right){}^*}{\sqrt{2}} $. Accordingly, the relations of  decay width between six channels can be deduced directly, with the effect of phase spaces ignored, given as
\begin{eqnarray*}
&&\Gamma(P_{\bar{d}su}^{++}\to \Xi_c^+ l^+\nu)= 2 \Gamma(P_{\{\bar{u}u,\bar{d}d\}\bar{s}}^{+}\to \Xi_c^0 l^+\nu )=2\Gamma(P_{\{\bar{d}d,\bar{s}s\}\bar{u}}^{++}\to \Lambda_c^+ l^+\nu),\\
&&\Gamma(P_{\bar{s}ud}^{++}\to \Lambda_c^+ l^+\nu)=2 \Gamma(P_{\{\bar{u}u,\bar{s}s\}\bar{d}}^{+}\to \Xi_c^0 l^+\nu)=2\Gamma(_{\{\bar{d}d,\bar{s}s\}\bar{u}}^{++}\to \Xi_c^+ l^+\nu).
\end{eqnarray*}

The transition of charmed quark decay $c \to q\bar{q}q$ can lead to non-leptonic decays of pentaquark $c c\bar qqq$ states $P_6$. Immediately, it is straightforward to construct the Hamiltonian in the hadronic level under the SU(3) light quark symmetry. We raise the possible Hamiltonian of sextet pentaquark $P_6$ decay into charmed mesons and light baryon $T_8$, $T_{10}$ as follows.
\begin{eqnarray}
\begin{split}
	   H&=a_1 (P_{6})_{\{ij\}} (H_{\overline{6}})_{k}^{[il]}(\bar D)^{j}(\overline{T}_{8})_{l}^{k}+a_2 (P_{6})_{\{ij\}}(H_{\overline{6}})_{k}^{[il]}(\bar D)^{k}(\overline{T}_{8})_{l}^{j}+a_3 (P_{6})_{\{ij\}} (H_{15})_{k}^{\{il\}}(\bar D)^{j}(\overline{T}_{8})_{l}^{k}\\
&+a_4 (P_{6})_{\{ij\}} (H_{15})_{k}^{\{il\}}(\bar D)^{k}(\overline{T}_{8})_{l}^{j}+a_5 (P_{6})_{\{ij\}} (H_{15})_{k}^{\{ij\}}(\bar D)^{l}(\overline{T}_{8})_{l}^{k},\\
       H&=b_1 (P_{6})_{\{ij\}} (H_{\overline{6}})_{k}^{[jl]}(\bar D)^{k}(\overline{T}_{10})_{\{l\alpha\beta\}}\varepsilon^{i\alpha\beta}+b_2 (P_{6})_{\{ij\}} (H_{\overline{6}})_{k}^{[jl]}(\bar D)^{\alpha}(\overline{T}_{10})_{\{l\alpha\beta\}}\varepsilon^{ik\beta}\\
&+b_3 (P_{6})_{\{ij\}} (H_{15})_{k}^{\{jl\}}(\bar D)^{k}(\overline{T}_{10})_{\{l\alpha\beta\}}\varepsilon^{i\alpha\beta}+b_4 (P_{6})_{\{ij\}} (H_{15})_{k}^{\{jl\}}(\bar D)^{\alpha}(\overline{T}_{10})_{\{l\alpha\beta\}}\varepsilon^{ik\beta}\\
&+b_5 (P_{6})_{\{ij\}} (H_{15})_{k}^{\{l\alpha\}}(\bar D)^{j}(\overline{T}_{10})_{\{l\alpha\beta\}}\varepsilon^{ik\beta}.
\end{split}
\end{eqnarray}
We expand the Hamiltonian and collect the possible processes, entrying into Tab.~\ref{tab:P6_DF8} and Tab.~\ref{tab:P6_DF8sc}., Meanwhile, it is ready to reduce the relations of decay width between different channels. We deduced the relations as follows.
\begin{eqnarray*}
&&\Gamma(P_{\bar{u}ds}^{0}\to  D^0 \Lambda^0)= 3\Gamma(P_{\bar{u}ds}^{0}\to  D^0 \Sigma^0),
 \Gamma(P_{\bar{d}su}^{++}\to  D^+ \Sigma^+)= \Gamma(P_{\bar{s}ud}^{++}\to  D^+_s p),\\
&&\Gamma(P_{\{\bar{u}u,\bar{s}s\}\bar{d}}^{+}\to  D^0 p)= \Gamma(P_{\{\bar{u}u,\bar{d}d\}\bar{s}}^{+}\to  D^0 \Sigma^+),
 \Gamma(P_{\{\bar{d}d,\bar{s}s\}\bar{u}}^{++}\to  D^+ p)= \Gamma(P_{\{\bar{d}d,\bar{s}s\}\bar{u}}^{++}\to  D^+_s \Sigma^+),\\
 &&\Gamma(P_{\bar{u}ds}^{0}\to  D^+ \Delta^{-})= 3\Gamma(P_{\bar{u}ds}^{0}\to  D^+_s \Sigma^{\prime-}),
 \Gamma(P_{\bar{u}ds}^{0}\to  D^+ \Sigma^{\prime-})= \Gamma(P_{\bar{u}ds}^{0}\to  D^+_s \Xi^{\prime-}),\\
 &&\Gamma(P_{\bar{u}ds}^{0}\to  D^+ \Xi^{\prime-})= \frac{1}{3}\Gamma(P_{\bar{u}ds}^{0}\to  D^+_s \Omega^-),
 \Gamma(P_{\bar{d}su}^{++}\to  D^0 \Delta^{++})= 3\Gamma(P_{\bar{d}su}^{++}\to  D^+_s \Sigma^{\prime+}),\\
 &&\Gamma(P_{\bar{d}su}^{++}\to  D^+ \Sigma^{\prime+})= \Gamma(P_{\bar{s}ud}^{++}\to  D^+_s \Delta^{+}),
 \Gamma(P_{\bar{s}ud}^{++}\to  D^0 \Delta^{++})= 3\Gamma(P_{\bar{s}ud}^{++}\to  D^+ \Delta^{+}),\\
 &&\Gamma(P_{\{\bar{u}u,\bar{s}s\}\bar{d}}^{+}\to  D^0 \Delta^{+})= \Gamma(P_{\{\bar{u}u,\bar{d}d\}\bar{s}}^{+}\to  D^0 \Sigma^{\prime+}),
 \Gamma(P_{\{\bar{u}u,\bar{s}s\}\bar{d}}^{+}\to  D^+ \Delta^{0})= \Gamma(P_{\{\bar{u}u,\bar{d}d\}\bar{s}}^{+}\to  D^+_s \Xi^{\prime0}),\\
 &&\Gamma(P_{\{\bar{u}u,\bar{s}s\}\bar{d}}^{+}\to  D^+_s \Delta^{0})= \Gamma(P_{\{\bar{d}d,\bar{s}s\}\bar{u}}^{++}\to  D^+_s \Delta^{+}),
 \Gamma(P_{\{\bar{u}u,\bar{s}s\}\bar{d}}^{+}\to  D^+_s \Sigma^{\prime0})= \Gamma(P_{\{\bar{u}u,\bar{d}d\}\bar{s}}^{+}\to  D^+ \Sigma^{\prime0}),\\
 &&\Gamma(P_{\{\bar{u}u,\bar{d}d\}\bar{s}}^{+}\to  D^+ \Xi^{\prime0})= \Gamma(P_{\{\bar{d}d,\bar{s}s\}\bar{u}}^{++}\to  D^+ \Sigma^{\prime+}),
 \Gamma(P_{\{\bar{d}d,\bar{s}s\}\bar{u}}^{++}\to  D^+ \Delta^{+})= \Gamma(P_{\{\bar{d}d,\bar{s}s\}\bar{u}}^{++}\to  D^+_s \Sigma^{\prime+}).
 \end{eqnarray*}
 \begin{table}
\caption{Pentaquark $P_{6}(cc\bar{q}qq)$ decays into charmed mesons and octet light baryons with  the Cabibbo allowed,
singly Cabibbo suppressed and doubly Cabibbo suppressed. sC represents the overall CKM factor $\sin(\theta_ C)$.}\label{tab:P6_DF8}\begin{tabular}{|c|c|c|c|c|c|c|c}\hline\hline
channel & amplitude&channel & amplitude \\\hline
$P_{\bar{s}ud}^{++}\to    D^+  p $ & $ \frac{a_2+a_4}{\sqrt{2}}$&
$P_{\bar{s}ud}^{++}\to    D^+_s  \Sigma^+ $ & $ \frac{a_1+a_3}{\sqrt{2}}$\\\hline
$P_{\{\bar{u}u,\bar{s}s\}\bar{d}}^{+}\to    D^0  \Sigma^+ $ & $ \frac{\left(a_1+a_3+2 a_5\right)}{2}$&
$P_{\{\bar{u}u,\bar{s}s\}\bar{d}}^{+}\to    D^+  \Lambda^0 $ & $ \frac{3 a_2-a_4+2 a_5}{2 \sqrt{6}}$\\\hline
$P_{\{\bar{u}u,\bar{s}s\}\bar{d}}^{+}\to    D^+  \Sigma^0 $ & $ \frac{a_2+a_4-2 a_5}{2 \sqrt{2}}$&
$P_{\{\bar{d}d,\bar{s}s\}\bar{u}}^{++}\to    D^+  \Sigma^+ $ & $ \frac{\left(a_1+a_2+a_3+a_4\right)}{2}$\\\hline
\hline$P_{\bar{u}ds}^{0}\to    D^0  \Lambda^0 $ & $ \frac{\sqrt{3}}{2}  \left(a_1-a_3\right) \text{sC}$&
$P_{\bar{u}ds}^{0}\to    D^0  \Sigma^0 $ & $ -\frac{1}{2} \left(a_1-a_3\right) \text{sC}$\\\hline
$P_{\bar{u}ds}^{0}\to    D^+  \Sigma^- $ & $ \frac{\left(a_2-a_4\right) \text{sC}}{\sqrt{2}}$&
$P_{\bar{d}su}^{++}\to    D^+  \Sigma^+ $ & $ -\frac{\left(a_1+a_2+a_3+a_4\right) \text{sC}}{\sqrt{2}}$\\\hline
$P_{\bar{s}ud}^{++}\to    D^+_s  p $ & $ \frac{\left(a_1+a_2+a_3+a_4\right) \text{sC}}{\sqrt{2}}$&
$P_{\{\bar{u}u,\bar{s}s\}\bar{d}}^{+}\to    D^0  p $ & $ \frac{\left(a_1+a_3+2 a_5\right) \text{sC}}{2}$\\\hline
$P_{\{\bar{u}u,\bar{s}s\}\bar{d}}^{+}\to    D^+  n $ & $ \frac{ \left(a_2-a_4+2 a_5\right) \text{sC}}{2}$&
$P_{\{\bar{u}u,\bar{s}s\}\bar{d}}^{+}\to    D^+_s  \Lambda^0 $ & $ \frac{\left(3 a_1+3 a_2-3 a_3-a_4-4 a_5\right) \text{sC}}{2 \sqrt{6}}$\\\hline
$P_{\{\bar{u}u,\bar{s}s\}\bar{d}}^{+}\to    D^+_s  \Sigma^0 $ & $ \frac{\left(-a_1+a_2+a_3+a_4\right) \text{sC}}{2 \sqrt{2}}$&
$P_{\{\bar{u}u,\bar{d}d\}\bar{s}}^{+}\to    D^0  \Sigma^+ $ & $ -\frac{ \left(a_1+a_3+2 a_5\right) \text{sC}}{2}$\\\hline
$P_{\{\bar{u}u,\bar{d}d\}\bar{s}}^{+}\to    D^+  \Lambda^0 $ & $ \frac{ \sqrt{6} \left(3  a_1-3 a_3-2  a_4-2  a_5\right)}{12} \text{sC}$&
$P_{\{\bar{u}u,\bar{d}d\}\bar{s}}^{+}\to    D^+  \Sigma^0 $ & $ -\frac{\left(a_1+2 a_2-a_3-2 a_5\right) \text{sC}}{2 \sqrt{2}}$\\\hline
$P_{\{\bar{d}d,\bar{s}s\}\bar{u}}^{++}\to    D^+  p $ & $ \frac{\left(a_1-a_2+a_3-a_4\right) \text{sC}}{2}$&
$P_{\{\bar{d}d,\bar{s}s\}\bar{u}}^{++}\to    D^+_s  \Sigma^+ $ & $ -\frac{ \left(a_1-a_2+a_3-a_4\right) \text{sC}}{2}$\\\hline
\hline$P_{\bar{u}ds}^{0}\to    D^0  n $ & $ -\frac{\left(a_1-a_3\right) \text{sC}^2}{\sqrt{2}}$&
$P_{\bar{u}ds}^{0}\to    D^+_s  \Sigma^- $ & $ \frac{\left(a_4-a_2\right) \text{sC}^2}{\sqrt{2}}$\\\hline
$P_{\bar{d}su}^{++}\to    D^+  p $ & $ \frac{\left(a_1+a_3\right) \text{sC}^2}{\sqrt{2}}$&
$P_{\bar{d}su}^{++}\to    D^+_s  \Sigma^+ $ & $ \frac{\left(a_2+a_4\right) \text{sC}^2}{\sqrt{2}}$\\\hline
$P_{\{\bar{u}u,\bar{s}s\}\bar{d}}^{+}\to    D^+_s  n $ & $ -\frac{ \left(a_1+a_2-a_3-a_4\right) \text{sC}^2}{2}$&
$P_{\{\bar{u}u,\bar{d}d\}\bar{s}}^{+}\to    D^0  p $ & $ \frac{ \left(a_1+a_3+2 a_5\right) \text{sC}^2}{2}$\\\hline
$P_{\{\bar{u}u,\bar{d}d\}\bar{s}}^{+}\to    D^+  n $ & $ -\frac{\left(a_1-a_3-2 a_5\right) \text{sC}^2}{2}$&
$P_{\{\bar{u}u,\bar{d}d\}\bar{s}}^{+}\to    D^+_s  \Lambda^0 $ & $ \frac{\left(a_4-2 a_5\right) \text{sC}^2}{\sqrt{6}}$\\\hline
$P_{\{\bar{u}u,\bar{d}d\}\bar{s}}^{+}\to    D^+_s  \Sigma^0 $ & $ \frac{a_2 \text{sC}^2}{\sqrt{2}}$&
$P_{\{\bar{d}d,\bar{s}s\}\bar{u}}^{++}\to    D^+_s  p $ & $ \frac{\left(a_1+a_2+a_3+a_4\right) \text{sC}^2}{2}$\\\hline
\hline
\end{tabular}
\end{table}
\begin{table}
\caption{Pentaquark $P_{6}(cc\bar{q}qq)$ decays into charmed mesons and decuplet light baryons with Cabibbo allowed, the
singly Cabibbo suppressed and doubly Cabibbo suppressed.}\label{tab:P6_DF8sc}\begin{tabular}{|c|c|c|c|c|c|c|c}\hline\hline
channel & amplitude&channel & amplitude \\\hline
$P_{\bar{u}ds}^{0}\to    D^0  \Xi^{\prime0} $ & $ \frac{-b_2+b_4+2 b_5}{\sqrt{6}}$&
$P_{\bar{u}ds}^{0}\to    D^+  \Xi^{\prime-} $ & $ \frac{b_4-b_2}{\sqrt{6}}$\\\hline
$P_{\bar{u}ds}^{0}\to    D^+_s  \Omega^- $ & $ \frac{b_4-b_2}{\sqrt{2}}$&
$P_{\bar{s}ud}^{++}\to    D^0  \Delta^{++} $ & $ -\frac{b_2+b_4}{\sqrt{2}}$\\\hline
$P_{\bar{s}ud}^{++}\to    D^+  \Delta^{+} $ & $ -\frac{b_2+b_4}{\sqrt{6}}$&
$P_{\bar{s}ud}^{++}\to    D^+_s  \Sigma^{\prime+} $ & $ -\frac{b_2+b_4+2 b_5}{\sqrt{6}}$\\\hline
$P_{\{\bar{u}u,\bar{s}s\}\bar{d}}^{+}\to    D^0  \Sigma^{\prime+} $ & $ \frac{b_2-b_5}{\sqrt{3}}$&
$P_{\{\bar{u}u,\bar{s}s\}\bar{d}}^{+}\to    D^+  \Sigma^{\prime0} $ & $ \frac{b_2}{\sqrt{6}}$\\\hline
$P_{\{\bar{u}u,\bar{s}s\}\bar{d}}^{+}\to    D^+_s  \Xi^{\prime0} $ & $ \frac{b_2+b_5}{\sqrt{3}}$&
$P_{\{\bar{u}u,\bar{d}d\}\bar{s}}^{+}\to    D^+  \Xi^{\prime0} $ & $ \frac{b_5}{\sqrt{3}}$\\\hline
$P_{\{\bar{d}d,\bar{s}s\}\bar{u}}^{++}\to    D^+  \Sigma^{\prime+} $ & $ -\frac{b_5}{\sqrt{3}}$&&\\\hline
\hline$P_{\bar{u}ds}^{0}\to    D^0  \Sigma^{\prime0} $ & $ \frac{\left(b_2-b_4-2 b_5\right) \text{sC}}{\sqrt{3}}$&
$P_{\bar{u}ds}^{0}\to    D^+  \Sigma^{\prime-} $ & $ \sqrt{\frac{2}{3}} \left(b_2-b_4\right) \text{sC}$\\\hline
$P_{\bar{u}ds}^{0}\to    D^+_s  \Xi^{\prime-} $ & $ \sqrt{\frac{2}{3}} \left(b_2-b_4\right) \text{sC}$&
$P_{\bar{d}su}^{++}\to    D^+  \Sigma^{\prime+} $ & $ \sqrt{\frac{2}{3}} b_5 \text{sC}$\\\hline
$P_{\bar{s}ud}^{++}\to    D^+_s  \Delta^{+} $ & $ \sqrt{\frac{2}{3}} b_5 \text{sC}$&
$P_{\{\bar{u}u,\bar{s}s\}\bar{d}}^{+}\to    D^0  \Delta^{+} $ & $ \frac{\left(b_5-b_2\right) \text{sC}}{\sqrt{3}}$\\\hline
$P_{\{\bar{u}u,\bar{s}s\}\bar{d}}^{+}\to    D^+  \Delta^{0} $ & $ -\frac{b_2 \text{sC}}{\sqrt{3}}$&
$P_{\{\bar{u}u,\bar{s}s\}\bar{d}}^{+}\to    D^+_s  \Sigma^{\prime0} $ & $ -\frac{\left(b_2+2 b_5\right) \text{sC}}{\sqrt{6}}$\\\hline
$P_{\{\bar{u}u,\bar{d}d\}\bar{s}}^{+}\to    D^0  \Sigma^{\prime+} $ & $ \frac{\left(b_5-b_2\right) \text{sC}}{\sqrt{3}}$&
$P_{\{\bar{u}u,\bar{d}d\}\bar{s}}^{+}\to    D^+  \Sigma^{\prime0} $ & $ -\frac{\left(b_2+2 b_5\right) \text{sC}}{\sqrt{6}}$\\\hline
$P_{\{\bar{u}u,\bar{d}d\}\bar{s}}^{+}\to    D^+_s  \Xi^{\prime0} $ & $ -\frac{b_2 \text{sC}}{\sqrt{3}}$&
$P_{\{\bar{d}d,\bar{s}s\}\bar{u}}^{++}\to    D^0  \Delta^{++} $ & $ \left(b_2+b_4\right) \text{sC}$\\\hline
$P_{\{\bar{d}d,\bar{s}s\}\bar{u}}^{++}\to    D^+  \Delta^{+} $ & $ \frac{\left(b_2+b_4+b_5\right) \text{sC}}{\sqrt{3}}$&
$P_{\{\bar{d}d,\bar{s}s\}\bar{u}}^{++}\to    D^+_s  \Sigma^{\prime+} $ & $ \frac{\left(b_2+b_4+b_5\right) \text{sC}}{\sqrt{3}}$\\\hline
\hline$P_{\bar{u}ds}^{0}\to    D^0  \Delta^{0} $ & $ \frac{\left(b_2-b_4-2 b_5\right) \text{sC}^2}{\sqrt{6}}$&
$P_{\bar{u}ds}^{0}\to    D^+  \Delta^{-} $ & $ \frac{\left(b_2-b_4\right) \text{sC}^2}{\sqrt{2}}$\\\hline
$P_{\bar{u}ds}^{0}\to    D^+_s  \Sigma^{\prime-} $ & $ \frac{\left(b_2-b_4\right) \text{sC}^2}{\sqrt{6}}$&
$P_{\bar{d}su}^{++}\to    D^0  \Delta^{++} $ & $ \frac{\left(b_2+b_4\right) \text{sC}^2}{\sqrt{2}}$\\\hline
$P_{\bar{d}su}^{++}\to    D^+  \Delta^{+} $ & $ \frac{\left( b_2+ b_4+2  b_5\right) \text{sC}^2}{\sqrt{6}}$&
$P_{\bar{d}su}^{++}\to    D^+_s  \Sigma^{\prime+} $ & $ \frac{\left(b_2+b_4\right) \text{sC}^2}{\sqrt{6}}$\\\hline
$P_{\{\bar{u}u,\bar{s}s\}\bar{d}}^{+}\to    D^+_s  \Delta^{0} $ & $ -\frac{b_5 \text{sC}^2}{\sqrt{3}}$&
$P_{\{\bar{u}u,\bar{d}d\}\bar{s}}^{+}\to    D^0  \Delta^{+} $ & $ \frac{\left(b_5-b_2\right) \text{sC}^2}{\sqrt{3}}$\\\hline
$P_{\{\bar{u}u,\bar{d}d\}\bar{s}}^{+}\to    D^+  \Delta^{0} $ & $ -\frac{\left(b_2+b_5\right) \text{sC}^2}{\sqrt{3}}$&
$P_{\{\bar{u}u,\bar{d}d\}\bar{s}}^{+}\to    D^+_s  \Sigma^{\prime0} $ & $ -\frac{b_2 \text{sC}^2}{\sqrt{6}}$\\\hline
$P_{\{\bar{d}d,\bar{s}s\}\bar{u}}^{++}\to    D^+_s  \Delta^{+} $ & $ \frac{b_5 \text{sC}^2}{\sqrt{3}}$&&\\\hline
\hline
\end{tabular}
\end{table}
More technically, the Hamiltonian of the sextet state $P_{6}$ decay into light mesons and singly charmed baryons can be constructed below.
\begin{eqnarray}
\begin{split}
		H&=c_1 (P_{6})_{\{ij\}}(H_{\bar6})_k^{[il]}(M)_{l}^j(\overline{T}_{c\bar3})^{k}+c_2 (P_{6})_{\{ij\}}(H_{\bar6})_k^{[il]}(M)_{l}^k(\overline{T}_{c\bar3})^{j}
           +c_3 (P_{6})_{\{ij\}}(H_{15})_k^{\{il\}}(M)_{l}^j(\overline{T}_{c\bar3})^{k}\\
&+c_4 (P_{6})_{\{ij\}}(H_{15})_k^{\{il\}}(M)_{l}^k(\overline{T}_{c\bar3})^{j}+c_5 (P_{6})_{\{ij\}}(H_{15})_k^{\{ij\}}(M)_{l}^k(\overline{T}_{c\bar3})^{l},\\
        H&=d_1 (P_{6})_{\{ij\}}(H_{\bar6})_k^{[jl]}(M)_{l}^{\alpha}(\overline{T}_{c6})_{\{\alpha\beta\}}\varepsilon^{ik\beta}+d_2 (P_{6})_{\{ij\}}(H_{\bar6})_k^{[j\alpha]}(M)_{l}^{k}(\overline{T}_{c6})_{\{\alpha\beta\}}\varepsilon^{il\beta}\\
&+d_3 (P_{6})_{\{ij\}}(H_{\bar6})_k^{[j\alpha]}(M)_{l}^{\beta}(\overline{T}_{c6})_{\{\alpha\beta\}}\varepsilon^{ikl}+d_4 (P_{6})_{\{ij\}}(H_{15})_k^{\{jl\}}(M)_{l}^{\alpha}(\overline{T}_{c6})_{\{\alpha\beta\}}\varepsilon^{ik\beta}\\
&+d_5 (P_{6})_{\{ij\}}(H_{15})_k^{\{j\alpha\}}(M)_{l}^{k}(\overline{T}_{c6})_{\{\alpha\beta\}}\varepsilon^{il\beta}+d_6 (P_{6})_{\{ij\}}(H_{15})_k^{\{j\alpha\}}(M)_{l}^{\beta}(\overline{T}_{c6})_{\{\alpha\beta\}}\varepsilon^{ikl}\\
&+d_7 (P_{6})_{\{ij\}}(H_{15})_k^{\{\alpha\beta\}}(M)_{l}^{j}(\overline{T}_{c6})_{\{\alpha\beta\}}\varepsilon^{ikl}.
\end{split}
\end{eqnarray}		
We expand the decay channels of pentaquark octet $P_{10}$ turning into anti-charmed mesons and light baryon, whose decay amplitudes collected in Tab.\ref{tab:P6_MF3} and Tab.\ref{tab:P6_MF3sc}. Moreover, the relations between different channels are given as follows.
\begin{eqnarray*}
&&\Gamma(P_{\bar{u}ds}^{0}\to \pi^0  \Xi_c^0)= \frac{1}{3}\Gamma(P_{\bar{u}ds}^{0}\to \eta_q  \Xi_c^0),
 \Gamma(P_{\bar{u}ds}^{0}\to \pi^-  \Xi_c^+)= \Gamma(P_{\bar{u}ds}^{0}\to K^-  \Lambda_c^+),\\
&&\Gamma(P_{\bar{d}su}^{++}\to \pi^+  \Xi_c^+)= \Gamma(P_{\bar{s}ud}^{++}\to K^+  \Lambda_c^+),
 \Gamma(P_{\{\bar{u}u,\bar{s}s\}\bar{d}}^{+}\to K^+  \Xi_c^0)= \Gamma(P_{\{\bar{u}u,\bar{d}d\}\bar{s}}^{+}\to \pi^+  \Xi_c^0),\\
&&\Gamma(P_{\{\bar{u}u,\bar{s}s\}\bar{d}}^{+}\to K^0  \Xi_c^+)= \Gamma(P_{\{\bar{u}u,\bar{d}d\}\bar{s}}^{+}\to \overline K^0  \Lambda_c^+),
 \Gamma(P_{\{\bar{d}d,\bar{s}s\}\bar{u}}^{++}\to \pi^+  \Lambda_c^+)=\Gamma(P_{\{\bar{d}d,\bar{s}s\}\bar{u}}^{++}\to K^+  \Xi_c^+),\\
&&\Gamma(P_{\bar{u}ds}^{0}\to \pi^0  \Xi_{c}^{\prime0})= 3\Gamma(P_{\bar{u}ds}^{0}\to \eta_q  \Xi_{c}^{\prime0}), \Gamma(P_{\{\bar{u}u,\bar{s}s\}\bar{d}}^{+}\to \pi^+  \Sigma_{c}^{0})= \Gamma(P_{\{\bar{u}u,\bar{d}d\}\bar{s}}^{+}\to K^+  \Omega_{c}^{0}),\\
&&\Gamma(P_{\bar{u}ds}^{0}\to \pi^-  \Xi_{c}^{\prime+})= \Gamma(P_{\bar{u}ds}^{0}\to K^-  \Sigma_{c}^{+}), \Gamma(P_{\bar{u}ds}^{0}\to K^0  \Omega_{c}^{0})= \Gamma(P_{\bar{u}ds}^{0}\to \overline K^0  \Sigma_{c}^{0}),\\
&&\Gamma(P_{\{\bar{u}u,\bar{s}s\}\bar{d}}^{+}\to K^+  \Xi_{c}^{\prime0})= \Gamma(P_{\{\bar{u}u,\bar{d}d\}\bar{s}}^{+}\to \pi^+  \Xi_{c}^{\prime0}),\Gamma(P_{\{\bar{u}u,\bar{s}s\}\bar{d}}^{+}\to K^0  \Xi_{c}^{\prime+})= \Gamma(P_{\{\bar{u}u,\bar{d}d\}\bar{s}}^{+}\to \overline K^0  \Sigma_{c}^{+}),\\
&&\Gamma(P_{\{\bar{d}d,\bar{s}s\}\bar{u}}^{++}\to \pi^+  \Xi_{c}^{\prime+})= \frac{1}{2}\Gamma(P_{\{\bar{d}d,\bar{s}s\}\bar{u}}^{++}\to \overline K^0  \Sigma_{c}^{++}),\Gamma(P_{\{\bar{d}d,\bar{s}s\}\bar{u}}^{++}\to K^+  \Sigma_{c}^{+})= \frac{1}{2}\Gamma(P_{\{\bar{d}d,\bar{s}s\}\bar{u}}^{++}\to K^0  \Sigma_{c}^{++}),\\
&&\Gamma(P_{\{\bar{u}u,\bar{s}s\}\bar{d}}^{+}\to \pi^-  \Sigma_{c}^{++})= \Gamma(P_{\{\bar{u}u,\bar{d}d\}\bar{s}}^{+}\to K^-  \Sigma_{c}^{++}), \Gamma(P_{\{\bar{u}u,\bar{s}s\}\bar{d}}^{+}\to K^+  \Sigma_{c}^{0})= 2\Gamma(P_{\{\bar{u}u,\bar{s}s\}\bar{d}}^{+}\to K^0  \Sigma_{c}^{+}),\\
&&\Gamma(P_{\bar{d}su}^{++}\to \pi^+  \Xi_{c}^{\prime+})= \frac{1}{2}\Gamma(P_{\bar{d}su}^{++}\to \overline K^0  \Sigma_{c}^{++})= \Gamma(P_{\bar{s}ud}^{++}\to K^+  \Sigma_{c}^{+})= \frac{1}{2}\Gamma(P_{\bar{s}ud}^{++}\to K^0  \Sigma_{c}^{++}),\\
&&\Gamma(P_{\{\bar{d}d,\bar{s}s\}\bar{u}}^{++}\to \pi^+  \Sigma_{c}^{+})= \Gamma(P_{\{\bar{d}d,\bar{s}s\}\bar{u}}^{++}\to K^+  \Xi_{c}^{\prime+})= 3\Gamma(P_{\{\bar{d}d,\bar{s}s\}\bar{u}}^{++}\to \eta_q  \Sigma_{c}^{++})= \Gamma(P_{\{\bar{d}d,\bar{s}s\}\bar{u}}^{++}\to \pi^0  \Sigma_{c}^{++}),\\
&&\Gamma(P_{\bar{u}ds}^{0}\to \pi^0  \Omega_{c}^{0})= \Gamma(P_{\{\bar{u}u,\bar{d}d\}\bar{s}}^{+}\to \pi^+  \Omega_{c}^{0})= 2\Gamma(P_{\{\bar{u}u,\bar{d}d\}\bar{s}}^{+}\to \overline K^0  \Xi_{c}^{\prime+}).
\end{eqnarray*}
\begin{table}
\caption{Pentaquark $P_{6}(cc\bar{q}qq)$ decays into light mesons and singly charmed baryons triplet with Cabibbo allowed, the
singly Cabibbo suppressed and doubly Cabibbo suppressed.}\label{tab:P6_MF3}\begin{tabular}{|c|c|c|c|c|c|c|c}\hline\hline
channel & amplitude&channel & amplitude \\\hline
$P_{\bar{u}ds}^{0}\to   \overline K^0   \Xi_c^0 $ & $ \frac{c_4-c_2}{\sqrt{2}}$&
$P_{\bar{u}ds}^{0}\to   K^-   \Xi_c^+ $ & $ \frac{c_1-c_3}{\sqrt{2}}$\\\hline
$P_{\bar{s}ud}^{++}\to   \pi^+   \Lambda_c^+ $ & $ \frac{c_2+c_4}{\sqrt{2}}$&
$P_{\bar{s}ud}^{++}\to   K^+   \Xi_c^+ $ & $ -\frac{c_1+c_3}{\sqrt{2}}$\\\hline
$P_{\{\bar{u}u,\bar{s}s\}\bar{d}}^{+}\to   \pi^+   \Xi_c^0 $ & $ \frac{1}{2} \left(c_2+c_4+2 c_5\right)$&
$P_{\{\bar{u}u,\bar{s}s\}\bar{d}}^{+}\to   \pi^0   \Xi_c^+ $ & $ -\frac{c_1+c_3-2 c_5}{2 \sqrt{2}}$\\\hline
$P_{\{\bar{u}u,\bar{s}s\}\bar{d}}^{+}\to   \overline K^0   \Lambda_c^+ $ & $ \frac{1}{2} \left(-c_2+c_4+2 c_5\right)$&
$P_{\{\bar{u}u,\bar{s}s\}\bar{d}}^{+}\to   \eta_q   \Xi_c^+ $ & $ \frac{-3 c_1+c_3-2 c_5}{2 \sqrt{6}}$\\\hline
$P_{\{\bar{u}u,\bar{d}d\}\bar{s}}^{+}\to   \overline K^0   \Xi_c^+ $ & $ \frac{1}{2} \left(c_1+c_2-c_3-c_4\right)$&
$P_{\{\bar{d}d,\bar{s}s\}\bar{u}}^{++}\to   \pi^+   \Xi_c^+ $ & $ \frac{\left(-c_1-c_2-c_3-c_4\right)}{2}$\\\hline
\hline$P_{\bar{u}ds}^{0}\to   \pi^0   \Xi_c^0 $ & $ -\frac{1}{2} \left(c_2-c_4\right) \text{sC}$&
$P_{\bar{u}ds}^{0}\to   \pi^-   \Xi_c^+ $ & $ \frac{\left(c_3-c_1\right) \text{sC}}{\sqrt{2}}$\\\hline
$P_{\bar{u}ds}^{0}\to   K^-   \Lambda_c^+ $ & $ \frac{\left(c_3-c_1\right) \text{sC}}{\sqrt{2}}$&
$P_{\bar{u}ds}^{0}\to   \eta_q   \Xi_c^0 $ & $ \frac{1}{2} \sqrt{3} \left(c_2-c_4\right) \text{sC}$\\\hline
$P_{\bar{d}su}^{++}\to   \pi^+   \Xi_c^+ $ & $ \frac{\left(c_1+c_2+c_3+c_4\right) \text{sC}}{\sqrt{2}}$&
$P_{\bar{s}ud}^{++}\to   K^+   \Lambda_c^+ $ & $ \frac{\left(c_1+c_2+c_3+c_4\right) \text{sC}}{\sqrt{2}}$\\\hline
$P_{\{\bar{u}u,\bar{s}s\}\bar{d}}^{+}\to   \pi^0   \Lambda_c^+ $ & $ \frac{\left(c_1-c_2+c_3+c_4\right) \text{sC}}{2 \sqrt{2}}$&
$P_{\{\bar{u}u,\bar{s}s\}\bar{d}}^{+}\to   K^+   \Xi_c^0 $ & $ \frac{1}{2} \left(c_2+c_4+2 c_5\right) \text{sC}$\\\hline
$P_{\{\bar{u}u,\bar{s}s\}\bar{d}}^{+}\to   K^0   \Xi_c^+ $ & $ -\frac{1}{2} \left(c_1-c_3+2 c_5\right) \text{sC}$&
$P_{\{\bar{u}u,\bar{s}s\}\bar{d}}^{+}\to   \eta_q   \Lambda_c^+ $ & $ \frac{ \sqrt{6}\left(3  c_1+3  c_2- c_3-3  c_4-4  c_5\right) \text{sC}}{12}$\\\hline
$P_{\{\bar{u}u,\bar{d}d\}\bar{s}}^{+}\to   \pi^+   \Xi_c^0 $ & $ -\frac{1}{2} \left(c_2+c_4+2 c_5\right) \text{sC}$&
$P_{\{\bar{u}u,\bar{d}d\}\bar{s}}^{+}\to   \pi^0   \Xi_c^+ $ & $ \frac{ \sqrt{2}\left(2  c_1+ c_2- c_4-2  c_5\right) \text{sC}}{4}$\\\hline
$P_{\{\bar{u}u,\bar{d}d\}\bar{s}}^{+}\to   \overline K^0   \Lambda_c^+ $ & $ -\frac{1}{2} \left(c_1-c_3+2 c_5\right) \text{sC}$&
$P_{\{\bar{u}u,\bar{d}d\}\bar{s}}^{+}\to   \eta_q   \Xi_c^+ $ & $ \frac{\left(-3 c_2+2 c_3+3 c_4+2 c_5\right) \text{sC}}{2 \sqrt{6}}$\\\hline
$P_{\{\bar{d}d,\bar{s}s\}\bar{u}}^{++}\to   \pi^+   \Lambda_c^+ $ & $ \frac{ \left(c_1-c_2+c_3-c_4\right) \text{sC}}{2}$&
$P_{\{\bar{d}d,\bar{s}s\}\bar{u}}^{++}\to   K^+   \Xi_c^+ $ & $ \frac{ \left(c_1-c_2+c_3-c_4\right) \text{sC}}{2}$\\\hline
\hline$P_{\bar{u}ds}^{0}\to   \pi^-   \Lambda_c^+ $ & $ -\frac{\left(c_1-c_3\right) \text{sC}^2}{\sqrt{2}}$&
$P_{\bar{u}ds}^{0}\to   K^0   \Xi_c^0 $ & $ \frac{\left(c_4-c_2\right) \text{sC}^2}{\sqrt{2}}$\\\hline
$P_{\bar{d}su}^{++}\to   \pi^+   \Lambda_c^+ $ & $ \frac{\left(c_1+c_3\right) \text{sC}^2}{\sqrt{2}}$&
$P_{\bar{d}su}^{++}\to   K^+   \Xi_c^+ $ & $ -\frac{1}{2} \left(\sqrt{2} c_2+\sqrt{2} c_4\right) \text{sC}^2$\\\hline
$P_{\{\bar{u}u,\bar{s}s\}\bar{d}}^{+}\to   K^0   \Lambda_c^+ $ & $ -\frac{\left(c_1+c_2-c_3-c_4\right) \text{sC}^2}{2}$&
$P_{\{\bar{u}u,\bar{d}d\}\bar{s}}^{+}\to   \pi^0   \Lambda_c^+ $ & $ \frac{c_1 \text{sC}^2}{\sqrt{2}}$\\\hline
$P_{\{\bar{u}u,\bar{d}d\}\bar{s}}^{+}\to   K^+   \Xi_c^0 $ & $ \frac{1}{2} \left(c_2+c_4+2 c_5\right) \text{sC}^2$&
$P_{\{\bar{u}u,\bar{d}d\}\bar{s}}^{+}\to   K^0   \Xi_c^+ $ & $ \frac{1}{2} \left(c_2-c_4-2 c_5\right) \text{sC}^2$\\\hline
$P_{\{\bar{u}u,\bar{d}d\}\bar{s}}^{+}\to   \eta_q   \Lambda_c^+ $ & $ \frac{\left(c_3-2 c_5\right) \text{sC}^2}{\sqrt{6}}$&
$P_{\{\bar{d}d,\bar{s}s\}\bar{u}}^{++}\to   K^+   \Lambda_c^+ $ & $ \frac{ \left(c_1+c_2+c_3+c_4\right) \text{sC}^2}{2}$\\\hline
\hline
\end{tabular}
\end{table}
\begin{table}
\caption{Pentaquark $P_{6}(cc\bar{q}qq)$ decays into light mesons and singly charmed baryons sextet with Cabibbo allowed, the
singly Cabibbo suppressed and doubly Cabibbo suppressed.}\label{tab:P6_MF3sc}\begin{tabular}{|c|c|c|c|c|c|c|c}\hline\hline
channel & amplitude&channel & amplitude \\\hline
$P_{\bar{u}ds}^{0}\to   \pi^0   \Omega_{c}^{0} $ & $ \frac{1}{2} \left(d_2-d_5\right)$&
$P_{\bar{u}ds}^{0}\to   \overline K^0   \Xi_{c}^{\prime0} $ & $ \frac{\left(-d_1+d_2-d_3+d_4-d_5+d_6\right)}{2}$\\\hline
$P_{\bar{u}ds}^{0}\to   K^-   \Xi_{c}^{\prime+} $ & $ \frac{ \left(-d_1-d_3+d_4+d_6\right)}{2}$&
$P_{\bar{u}ds}^{0}\to   \eta_q   \Omega_{c}^{0} $ & $ \frac{2 d_1-d_2+2 d_3-2 d_4+d_5-2 d_6}{2 \sqrt{3}}$\\\hline
$P_{\bar{s}ud}^{++}\to   \pi^+   \Sigma_{c}^{+} $ & $ \frac{ \left(-d_1+d_2-d_3-d_4+d_5-d_6\right)}{2}$&
$P_{\bar{s}ud}^{++}\to   \pi^0   \Sigma_{c}^{++} $ & $ \frac{ \left(-d_1+d_2-d_3-d_4+d_5-d_6\right)}{2}$\\\hline
$P_{\bar{s}ud}^{++}\to   K^+   \Xi_{c}^{\prime+} $ & $ \frac{ \left(-d_1-d_3-d_4-d_6\right)}{2}$&
$P_{\bar{s}ud}^{++}\to   \eta_q   \Sigma_{c}^{++} $ & $ -\frac{d_1+d_2+d_3+d_4+d_5+d_6}{2 \sqrt{3}}$\\\hline
$P_{\{\bar{u}u,\bar{s}s\}\bar{d}}^{+}\to   \pi^+   \Xi_{c}^{\prime0} $ & $ \frac{d_1-d_2+d_3+d_4+d_5-d_6}{2 \sqrt{2}}$&
$P_{\{\bar{u}u,\bar{s}s\}\bar{d}}^{+}\to   \pi^0   \Xi_{c}^{\prime+} $ & $ \frac{ \left(d_1-2 d_2+d_3+d_4-d_6\right)}{4}$\\\hline
$P_{\{\bar{u}u,\bar{s}s\}\bar{d}}^{+}\to   K^+   \Omega_{c}^{0} $ & $ \frac{1}{2} \left(d_1+d_3+d_4-d_6\right)$&
$P_{\{\bar{u}u,\bar{s}s\}\bar{d}}^{+}\to   \overline K^0   \Sigma_{c}^{+} $ & $ \frac{d_1-d_2+d_3-d_4-d_5+d_6}{2 \sqrt{2}}$\\\hline
$P_{\{\bar{u}u,\bar{s}s\}\bar{d}}^{+}\to   K^-   \Sigma_{c}^{++} $ & $ \frac{1}{2} \left(d_1+d_3-d_4+d_6\right)$&
$P_{\{\bar{u}u,\bar{s}s\}\bar{d}}^{+}\to   \eta_q   \Xi_{c}^{\prime+} $ & $ -\frac{d_1-2 d_2+d_3-3 d_4+3 d_6}{4 \sqrt{3}}$\\\hline
$P_{\{\bar{u}u,\bar{d}d\}\bar{s}}^{+}\to   \pi^+   \Omega_{c}^{0} $ & $ \frac{1}{2} \left(d_2-d_5\right)$&
$P_{\{\bar{u}u,\bar{d}d\}\bar{s}}^{+}\to   \overline K^0   \Xi_{c}^{\prime+} $ & $ \frac{d_5-d_2}{2 \sqrt{2}}$\\\hline
$P_{\{\bar{d}d,\bar{s}s\}\bar{u}}^{++}\to   \pi^+   \Xi_{c}^{\prime+} $ & $ -\frac{d_2+d_5}{2 \sqrt{2}}$&
$P_{\{\bar{d}d,\bar{s}s\}\bar{u}}^{++}\to   \overline K^0   \Sigma_{c}^{++} $ & $ \frac{1}{2} \left(d_2+d_5\right)$\\\hline
\hline$P_{\bar{u}ds}^{0}\to   \pi^0   \Xi_{c}^{\prime0} $ & $ -\frac{\left(d_1+d_2+d_3-d_4-d_5-d_6\right) \text{sC}}{2 \sqrt{2}}$&
$P_{\bar{u}ds}^{0}\to   \pi^-   \Xi_{c}^{\prime+} $ & $ \frac{1}{2} \left(d_1+d_3-d_4-d_6\right) \text{sC}$\\\hline
$P_{\bar{u}ds}^{0}\to   K^0   \Omega_{c}^{0} $ & $ \frac{\left(d_1-d_2+d_3-d_4+d_5-d_6\right) \text{sC}}{\sqrt{2}}$&
$P_{\bar{u}ds}^{0}\to   \overline K^0   \Sigma_{c}^{0} $ & $ \frac{\left(d_1-d_2+d_3-d_4+d_5-d_6\right) \text{sC}}{\sqrt{2}}$\\\hline
$P_{\bar{u}ds}^{0}\to   K^-   \Sigma_{c}^{+} $ & $ \frac{1}{2} \left(d_1+d_3-d_4-d_6\right) \text{sC}$&
$P_{\bar{u}ds}^{0}\to   \eta_q   \Xi_{c}^{\prime0} $ & $ \frac{\left(-d_1-d_2-d_3+d_4+d_5+d_6\right) \text{sC}}{2 \sqrt{6}}$\\\hline
$P_{\bar{d}su}^{++}\to   \pi^+   \Xi_{c}^{\prime+} $ & $ \frac{1}{2} \left(d_2+d_5\right) \text{sC}$&
$P_{\bar{d}su}^{++}\to   \overline K^0   \Sigma_{c}^{++} $ & $ -\frac{\left(d_2+d_5\right) \text{sC}}{\sqrt{2}}$\\\hline
$P_{\bar{s}ud}^{++}\to   K^+   \Sigma_{c}^{+} $ & $ \frac{1}{2} \left(d_2+d_5\right) \text{sC}$&
$P_{\bar{s}ud}^{++}\to   K^0   \Sigma_{c}^{++} $ & $ -\frac{\left(d_2+d_5\right) \text{sC}}{\sqrt{2}}$\\\hline
$P_{\{\bar{u}u,\bar{s}s\}\bar{d}}^{+}\to   \pi^+   \Sigma_{c}^{0} $ & $ -\frac{ \left(d_1-d_2+d_3+d_4+d_5-d_6\right) \text{sC}}{2}$&
$P_{\{\bar{u}u,\bar{s}s\}\bar{d}}^{+}\to   \pi^0   \Sigma_{c}^{+} $ & $ \frac{ \left(d_2-2 d_4-d_5+2 d_6\right) \text{sC}}{4}$\\\hline
$P_{\{\bar{u}u,\bar{s}s\}\bar{d}}^{+}\to   \pi^-   \Sigma_{c}^{++} $ & $ -\frac{1}{2} \left(d_1+d_3-d_4+d_6\right) \text{sC}$&
$P_{\{\bar{u}u,\bar{s}s\}\bar{d}}^{+}\to   K^+   \Xi_{c}^{\prime0} $ & $ -\frac{\left(d_1+d_2+d_3+d_4-d_5-d_6\right) \text{sC}}{2 \sqrt{2}}$\\\hline
$P_{\{\bar{u}u,\bar{s}s\}\bar{d}}^{+}\to   K^0   \Xi_{c}^{\prime+} $ & $ -\frac{\left(d_1-2 d_2+d_3-d_4+d_6\right) \text{sC}}{2 \sqrt{2}}$&
$P_{\{\bar{u}u,\bar{s}s\}\bar{d}}^{+}\to   \eta_q   \Sigma_{c}^{+} $ & $ \frac{\left(-2 d_1+d_2-2 d_3+3 d_5\right) \text{sC}}{4 \sqrt{3}}$\\\hline
$P_{\{\bar{u}u,\bar{d}d\}\bar{s}}^{+}\to   \pi^+   \Xi_{c}^{\prime0} $ & $ -\frac{\left(d_1+d_2+d_3+d_4-d_5-d_6\right) \text{sC}}{2 \sqrt{2}}$&
$P_{\{\bar{u}u,\bar{d}d\}\bar{s}}^{+}\to   \pi^0   \Xi_{c}^{\prime+} $ & $ -\frac{ \left(d_1-d_2+d_3+d_4-d_5-d_6\right) \text{sC}}{4}$\\\hline
$P_{\{\bar{u}u,\bar{d}d\}\bar{s}}^{+}\to   K^+   \Omega_{c}^{0} $ & $ -\frac{\left(d_1-d_2+d_3+d_4+d_5-d_6\right) \text{sC}}{2}$&
$P_{\{\bar{u}u,\bar{d}d\}\bar{s}}^{+}\to   \overline K^0   \Sigma_{c}^{+} $ & $ -\frac{\left(d_1-2 d_2+d_3-d_4+d_6\right) \text{sC}}{2 \sqrt{2}}$\\\hline
$P_{\{\bar{u}u,\bar{d}d\}\bar{s}}^{+}\to   K^-   \Sigma_{c}^{++} $ & $ -\frac{ \left(d_1+d_3-d_4+d_6\right) \text{sC}}{2}$&
$P_{\{\bar{u}u,\bar{d}d\}\bar{s}}^{+}\to   \eta_q   \Xi_{c}^{\prime+} $ & $ \frac{\left(d_1+d_2+d_3-3 d_4-3 d_5+3 d_6\right) \text{sC}}{4 \sqrt{3}}$\\\hline
$P_{\{\bar{d}d,\bar{s}s\}\bar{u}}^{++}\to   \pi^+   \Sigma_{c}^{+} $ & $ \frac{\left(2 d_1-d_2+2 d_3+2 d_4-d_5+2 d_6\right) \text{sC}}{2 \sqrt{2}}$&
$P_{\{\bar{d}d,\bar{s}s\}\bar{u}}^{++}\to   \pi^0   \Sigma_{c}^{++} $ & $ \frac{\left(2 d_1-d_2+2 d_3+2 d_4-d_5+2 d_6\right) \text{sC}}{2 \sqrt{2}}$\\\hline
$P_{\{\bar{d}d,\bar{s}s\}\bar{u}}^{++}\to   K^+   \Xi_{c}^{\prime+} $ & $ \frac{\left(2 d_1-d_2+2 d_3+2 d_4-d_5+2 d_6\right) \text{sC}}{2 \sqrt{2}}$&
$P_{\{\bar{d}d,\bar{s}s\}\bar{u}}^{++}\to   \eta_q   \Sigma_{c}^{++} $ & $ \frac{\left(2 d_1-d_2+2 d_3+2 d_4-d_5+2 d_6\right) \text{sC}}{2 \sqrt{6}}$\\\hline
\hline$P_{\bar{u}ds}^{0}\to   \pi^0   \Sigma_{c}^{0} $ & $ -\frac{ \left(d_1+d_3-d_4-d_6\right) \text{sC}^2}{2}$&
$P_{\bar{u}ds}^{0}\to   \pi^-   \Sigma_{c}^{+} $ & $ \frac{\left(d_1+d_3-d_4-d_6\right) \text{sC}^2}{2}$\\\hline
$P_{\bar{u}ds}^{0}\to   K^0   \Xi_{c}^{\prime0} $ & $ \frac{\left(d_1-d_2+d_3-d_4+d_5-d_6\right) \text{sC}^2}{2}$&
$P_{\bar{u}ds}^{0}\to   \eta_q   \Sigma_{c}^{0} $ & $ \frac{ \sqrt{3}\left( d_1-2  d_2+ d_3- d_4+2 d_5-d_6\right) \text{sC}^2}{6}$\\\hline
$P_{\bar{d}su}^{++}\to   \pi^+   \Sigma_{c}^{+} $ & $ \frac{ \left(d_1+d_3+d_4+d_6\right) \text{sC}^2}{2}$&
$P_{\bar{d}su}^{++}\to   \pi^0   \Sigma_{c}^{++} $ & $ \frac{1}{2} \left(d_1+d_3+d_4+d_6\right) \text{sC}^2$\\\hline
$P_{\bar{d}su}^{++}\to   K^+   \Xi_{c}^{\prime+} $ & $ \frac{\left(d_1-d_2+d_3+d_4-d_5+d_6\right) \text{sC}^2}{2}$&
$P_{\bar{d}su}^{++}\to   \eta_q   \Sigma_{c}^{++} $ & $ \frac{ \sqrt{3}\left( d_1-2 d_2+ d_3+ d_4-2  d_5+d_6\right) \text{sC}^2}{6}$\\\hline
$P_{\{\bar{u}u,\bar{s}s\}\bar{d}}^{+}\to   K^+   \Sigma_{c}^{0} $ & $ -\frac{1}{2} \left(d_2-d_5\right) \text{sC}^2$&
$P_{\{\bar{u}u,\bar{s}s\}\bar{d}}^{+}\to   K^0   \Sigma_{c}^{+} $ & $ \frac{\left(d_2-d_5\right) \text{sC}^2}{2 \sqrt{2}}$\\\hline
$P_{\{\bar{u}u,\bar{d}d\}\bar{s}}^{+}\to   \pi^+   \Sigma_{c}^{0} $ & $ -\frac{1}{2} \left(d_1+d_3+d_4-d_6\right) \text{sC}^2$&
$P_{\{\bar{u}u,\bar{d}d\}\bar{s}}^{+}\to   \pi^0   \Sigma_{c}^{+} $ & $ -\frac{1}{2} \left(d_4-d_6\right) \text{sC}^2$\\\hline
$P_{\{\bar{u}u,\bar{d}d\}\bar{s}}^{+}\to   \pi^-   \Sigma_{c}^{++} $ & $ -\frac{1}{2} \left(d_1+d_3-d_4+d_6\right) \text{sC}^2$&
$P_{\{\bar{u}u,\bar{d}d\}\bar{s}}^{+}\to   K^+   \Xi_{c}^{\prime0} $ & $ -\frac{\left(d_1-d_2+d_3+d_4+d_5-d_6\right) \text{sC}^2}{2 \sqrt{2}}$\\\hline
$P_{\{\bar{u}u,\bar{d}d\}\bar{s}}^{+}\to   K^0   \Xi_{c}^{\prime+} $ & $ -\frac{\left(d_1-d_2+d_3-d_4-d_5+d_6\right) \text{sC}^2}{2 \sqrt{2}}$&
$P_{\{\bar{u}u,\bar{d}d\}\bar{s}}^{+}\to   \eta_q   \Sigma_{c}^{+} $ & $ -\frac{ \sqrt{3}\left( d_1-2  d_2+ d_3\right) \text{sC}^2}{6}$\\\hline
$P_{\{\bar{d}d,\bar{s}s\}\bar{u}}^{++}\to   K^+   \Sigma_{c}^{+} $ & $ \frac{\left(d_2+d_5\right) \text{sC}^2}{2 \sqrt{2}}$&
$P_{\{\bar{d}d,\bar{s}s\}\bar{u}}^{++}\to   K^0   \Sigma_{c}^{++} $ & $ -\frac{1}{2} \left(d_2+d_5\right) \text{sC}^2$\\\hline
\hline
\end{tabular}
\end{table}

\begin{figure}
  \centering
  \includegraphics[width=0.95\columnwidth]{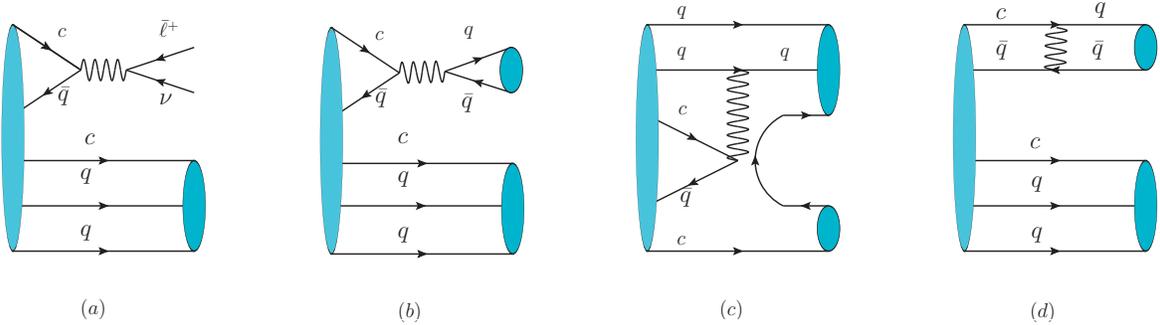}\\
  \caption{The weak decay diagrams of the doubly heavy pentaquark $P_{6}$ are given in (a,b,c,d), which semi-leptonic decays to be (a), non-leptonic decays to be (b,c,d).}\label{fig:topology1}
\end{figure}
\section{Golden Channels}
\label{sec:golden_channels}
As a collection, we will screen out some golden channels to produce and reconstruct the pentaquark $P_{cc\bar qqq}$ in the section. In principle, the main considerations are the CKM elements in the transition. The amplitudes of c-quark decay transitions such as  $c\to  s\bar d  u$ and $c\to  s\ell^+ \nu_{\ell}$ will receive the largest contribution as $V_{cs}^*\sim 1$. Beyond that, the detection efficiency is also a serviceable factor, generally speaking, charged particles have higher rates to be detected than neutral particles. According to the selection schemes, one use the following criteria~\cite{Xing:2019wil}, finally, we can obtain the golden decay channels in Table~\ref{tab:P6_golden_meson}.
\begin{itemize}
\item Branching fractions:  one chooses the corresponding channels with the quark transition of $\bar c\to \bar sd \bar u$ or $\bar c\to \bar s\ell^-   \bar \nu_{\ell}$.

\item Detection efficiency: one removes all channels with the hadrons $\pi^0$, $K^0$, $\overline{K}^0$, $n$, $\Sigma^{+}(\to p \pi^0)$ and $\Sigma^-(\to n \pi^-)$ in final states, but keep the processes with $\pi^\pm$, $\Sigma^{0}(\to N \pi \gamma)$, $\Lambda^0(\to p \pi^-)$ and $\Delta^0(\to N \pi)$.

\end{itemize}
\begin{table}
 \caption{The golden channels of production and decays with Cabibbo allowed $P(cc\bar qqq)$. }\label{tab:P6_golden_meson}\begin{tabular}{|c    c   c c|}\hline\hline
$\Omega_{ccc}^{++}\to   P_{\{\bar{u}u,\bar{d}d\}\bar{s}}^{+}  \pi^+$&
$\Omega_{ccc}^{++}\to   {P'}_{\eta s}^{+}  \pi^+  $&
$\Omega_{ccc}^{++}\to   {P'}_{\pi s}^{+}  \pi^+  $ &\\
$\Omega_{ccc}^{++}\to   {P'}_{ss\bar{d}}^{+}  K^+  $&&&\\
\hline\hline
$P_{\bar{s}ud}^{++}\to    D^+  p $&
$P_{\{\bar{u}u,\bar{s}s\}\bar{d}}^{+}\to    D^+  \Lambda^0 $&
$P_{\{\bar{u}u,\bar{s}s\}\bar{d}}^{+}\to    D^+  \Sigma^0 $&\\
$P_{\bar{u}ds}^{0}\to    D^+  \Xi^{\prime-} $&
$P_{\bar{u}ds}^{0}\to    D^+_s  \Omega^- $&
$P_{\bar{s}ud}^{++}\to    D^+  \Delta^{+} $&\\
$P_{\bar{s}ud}^{++}\to    D^+_s  \Sigma^{\prime+} $&
$P_{\{\bar{d}d,\bar{s}s\}\bar{u}}^{++}\to    D^+  \Sigma^{\prime+} $&
$P_{\bar{u}ds}^{0}\to   K^-   \Xi_c^+ $&\\
$P_{\bar{s}ud}^{++}\to   \pi^+   \Lambda_c^+ $&
$P_{\bar{s}ud}^{++}\to   K^+   \Xi_c^+ $&
$P_{\{\bar{d}d,\bar{s}s\}\bar{u}}^{++}\to   \pi^+   \Xi_c^+ $&\\
$P_{\bar{u}ds}^{0}\to   K^-   \Xi_{c}^{\prime+} $&
$P_{\bar{s}ud}^{++}\to   \pi^+   \Sigma_{c}^{+} $&
$P_{\bar{s}ud}^{++}\to   K^+   \Xi_{c}^{\prime+} $&\\
$P_{\{\bar{u}u,\bar{s}s\}\bar{d}}^{+}\to   K^-   \Sigma_{c}^{++} $&
$P_{\{\bar{d}d,\bar{s}s\}\bar{u}}^{++}\to   \pi^+   \Xi_{c}^{\prime+} $&&\\

\hline
\end{tabular}
\end{table}

\section{Conclusions}
\label{sec:conclusions}
In the paper, we discussed the mass spectrums of doubly charmed pentaquarks $P_{cc\bar qqq}$ primarily under the doubly heavy triquark-diquark model. Moreover, the lifetimes were considered at the next-to-leading order with the approach of OPE. The calculation suggested some potential stable pentaquark states against strong interaction, for instance, $cc\bar s ud$ with the parity $J^P=\frac{1}{2}^-$. Further more, the lifetimes obtained as $\tau(P_{cc\bar qqq}(\frac{1}{2}^-))=(4.65^{+0.71}_{-0.55})\times 10^{-13}s $ and $\tau(P_{cc\bar qqq}(\frac{3}{2}^-))=(0.93^{+0.14}_{-0.11})\times 10^{-12}  s $. Subsequently, we systematically discussed the production and decay behaviors of $P(cc\bar qqq)$, such as semi- or non-leptonic processes. Finally, we collected the golden channels of production and decay processes, which with the largest branching fraction and experimental detector efficiency. Our results are helpful to search for the  doubly charmed pentaquark $P_{cc\bar qqq}$ in future experiments.

\section*{Acknowledgments}

 This work is
supported in part by  National Natural Science Foundation of
China under Grant No.~12005294 and 11774417.


\begin{references}
\bibitem{Aaij:2015tga}
R.~Aaij \textit{et al.} [LHCb],
Phys. Rev. Lett. \textbf{115} (2015), 072001
doi:10.1103/PhysRevLett.115.072001
[arXiv:1507.03414 [hep-ex]].

\bibitem{Aaij:2016phn}
R.~Aaij \textit{et al.} [LHCb],
Phys. Rev. Lett. \textbf{117} (2016) no.8, 082002
doi:10.1103/PhysRevLett.117.082002
[arXiv:1604.05708 [hep-ex]].

\bibitem{Liu:2019tjn}
M.~Z.~Liu, Y.~W.~Pan, F.~Z.~Peng, M.~S\'anchez S\'anchez, L.~S.~Geng, A.~Hosaka and M.~Pavon Valderrama,
Phys. Rev. Lett. \textbf{122} (2019) no.24, 242001
doi:10.1103/PhysRevLett.122.242001
[arXiv:1903.11560 [hep-ph]].

\bibitem{He:2019ify}
J.~He,
Eur. Phys. J. C \textbf{79} (2019) no.5, 393
doi:10.1140/epjc/s10052-019-6906-1
[arXiv:1903.11872 [hep-ph]].

\bibitem{Meng:2019ilv}
L.~Meng, B.~Wang, G.~J.~Wang and S.~L.~Zhu,
Phys. Rev. D \textbf{100} (2019) no.1, 014031
doi:10.1103/PhysRevD.100.014031
[arXiv:1905.04113 [hep-ph]].

\bibitem{Azizi:2018bdv}
K.~Azizi, Y.~Sarac and H.~Sundu,
Phys. Lett. B \textbf{782} (2018), 694-701
doi:10.1016/j.physletb.2018.06.022
[arXiv:1802.01384 [hep-ph]].

\bibitem{Chen:2015moa}
H.~X.~Chen, W.~Chen, X.~Liu, T.~G.~Steele and S.~L.~Zhu,
Phys. Rev. Lett. \textbf{115} (2015) no.17, 172001
doi:10.1103/PhysRevLett.115.172001
[arXiv:1507.03717 [hep-ph]].

\bibitem{Richard:2019fms}
J.~M.~Richard, A.~Valcarce and J.~Vijande,
Phys. Lett. B \textbf{790} (2019), 248-250
doi:10.1016/j.physletb.2019.01.031
[arXiv:1901.03578 [hep-ph]].

\bibitem{Yang:2015bmv}
G.~Yang and J.~Ping,
Phys. Rev. D \textbf{95} (2017) no.1, 014010
doi:10.1103/PhysRevD.95.014010
[arXiv:1511.09053 [hep-ph]].

\bibitem{Huang:2015uda}
H.~Huang, C.~Deng, J.~Ping and F.~Wang,
Eur. Phys. J. C \textbf{76} (2016) no.11, 624
doi:10.1140/epjc/s10052-016-4476-z
[arXiv:1510.04648 [hep-ph]].

\bibitem{Ali:2019npk}
A.~Ali and A.~Y.~Parkhomenko,
Phys. Lett. B \textbf{793} (2019), 365-371
doi:10.1016/j.physletb.2019.05.002
[arXiv:1904.00446 [hep-ph]].

\bibitem{Ali:2019clg}
A.~Ali, I.~Ahmed, M.~J.~Aslam, A.~Y.~Parkhomenko and A.~Rehman,
JHEP \textbf{10} (2019), 256
doi:10.1007/JHEP10(2019)256
[arXiv:1907.06507 [hep-ph]].

\bibitem{Maiani:2014aja}
L.~Maiani, F.~Piccinini, A.~D.~Polosa and V.~Riquer,
Phys. Rev. D \textbf{89} (2014), 114010
doi:10.1103/PhysRevD.89.114010
[arXiv:1405.1551 [hep-ph]].

\bibitem{Lenz:2014jha}
A.~Lenz,
Int. J. Mod. Phys. A \textbf{30} (2015) no.10, 1543005
doi:10.1142/S0217751X15430058
[arXiv:1405.3601 [hep-ph]].

\bibitem{Ali:2018xfq}
A.~Ali, Q.~Qin and W.~Wang,
Phys. Lett. B \textbf{785} (2018), 605-609
doi:10.1016/j.physletb.2018.09.018
[arXiv:1806.09288 [hep-ph]].

\bibitem{Savage:1989ub}
M.~J.~Savage and M.~B.~Wise,
Phys. Rev. D \textbf{39} (1989), 3346
[erratum: Phys. Rev. D \textbf{40} (1989), 3127]
doi:10.1103/PhysRevD.39.3346

\bibitem{Gronau:1995hm}
M.~Gronau, O.~F.~Hernandez, D.~London and J.~L.~Rosner,
Phys. Rev. D \textbf{52} (1995), 6356-6373
doi:10.1103/PhysRevD.52.6356
[arXiv:hep-ph/9504326 [hep-ph]].

\bibitem{He:1998rq}
X.~G.~He,
Eur. Phys. J. C \textbf{9} (1999), 443-448
doi:10.1007/s100529900064
[arXiv:hep-ph/9810397 [hep-ph]].

\bibitem{Chiang:2004nm}
C.~W.~Chiang, M.~Gronau, J.~L.~Rosner and D.~A.~Suprun,
Phys. Rev. D \textbf{70} (2004), 034020
doi:10.1103/PhysRevD.70.034020
[arXiv:hep-ph/0404073 [hep-ph]].

\bibitem{Li:2007bh}
Y.~Li, C.~D.~Lu and W.~Wang,
Phys. Rev. D \textbf{77} (2008), 054001
doi:10.1103/PhysRevD.77.054001
[arXiv:0711.0497 [hep-ph]].

\bibitem{Wang:2009azc}
W.~Wang and C.~D.~Lu,
Phys. Rev. D \textbf{82} (2010), 034016
doi:10.1103/PhysRevD.82.034016
[arXiv:0910.0613 [hep-ph]].

\bibitem{Cheng:2011qh}
H.~Y.~Cheng and S.~Oh,
JHEP \textbf{09} (2011), 024
doi:10.1007/JHEP09(2011)024
[arXiv:1104.4144 [hep-ph]].

\bibitem{Hsiao:2015iiu}
Y.~K.~Hsiao, C.~F.~Chang and X.~G.~He,
Phys. Rev. D \textbf{93} (2016) no.11, 114002
doi:10.1103/PhysRevD.93.114002
[arXiv:1512.09223 [hep-ph]].

\bibitem{Lu:2016ogy}
C.~D.~L\"u, W.~Wang and F.~S.~Yu,
Phys. Rev. D \textbf{93} (2016) no.5, 056008
doi:10.1103/PhysRevD.93.056008
[arXiv:1601.04241 [hep-ph]].

\bibitem{He:2016xvd}
X.~G.~He, W.~Wang and R.~L.~Zhu,
J. Phys. G \textbf{44} (2017) no.1, 014003
doi:10.1088/0954-3899/44/1/014003
[arXiv:1606.00097 [hep-ph]].

\bibitem{Wang:2017vnc}
W.~Wang and R.~L.~Zhu,
Phys. Rev. D \textbf{96} (2017) no.1, 014024
doi:10.1103/PhysRevD.96.014024
[arXiv:1704.00179 [hep-ph]].

\bibitem{Wang:2017azm}
W.~Wang, Z.~P.~Xing and J.~Xu,
Eur. Phys. J. C \textbf{77} (2017) no.11, 800
doi:10.1140/epjc/s10052-017-5363-y
[arXiv:1707.06570 [hep-ph]].

\bibitem{Shi:2017dto}
Y.~J.~Shi, W.~Wang, Y.~Xing and J.~Xu,
Eur. Phys. J. C \textbf{78} (2018) no.1, 56
doi:10.1140/epjc/s10052-018-5532-7
[arXiv:1712.03830 [hep-ph]].

\bibitem{He:2018php}
X.~G.~He and W.~Wang,
Chin. Phys. C \textbf{42} (2018) no.10, 103108
doi:10.1088/1674-1137/42/10/103108
[arXiv:1803.04227 [hep-ph]].

\bibitem{Shi:2020gfp}
Y.~J.~Shi, Y.~Xing and Z.~X.~Zhao,
Eur. Phys. J. C \textbf{81} (2021) no.2, 156
doi:10.1140/epjc/s10052-021-08954-8
[arXiv:2012.12613 [hep-ph]].

\bibitem{Li:2021rfj}
D.~M.~Li, X.~R.~Zhang, Y.~Xing and J.~Xu,
[arXiv:2101.12574 [hep-ph]].

\bibitem{Jaffe:2003sg}
R.~L.~Jaffe and F.~Wilczek,
Phys. Rev. Lett. \textbf{91} (2003), 232003
doi:10.1103/PhysRevLett.91.232003
[arXiv:hep-ph/0307341 [hep-ph]].

\bibitem{Ali:2009pi}
A.~Ali, C.~Hambrock, I.~Ahmed and M.~J.~Aslam,
Phys. Lett. B \textbf{684} (2010), 28-39
doi:10.1016/j.physletb.2009.12.053
[arXiv:0911.2787 [hep-ph]].

\bibitem{Guo:2017vcf}
Z.~H.~Guo,
Phys. Rev. D \textbf{96} (2017) no.7, 074004
doi:10.1103/PhysRevD.96.074004
[arXiv:1708.04145 [hep-ph]].

\bibitem{Park:2018oib}
W.~Park, S.~Cho and S.~H.~Lee,
Phys. Rev. D \textbf{99} (2019) no.9, 094023
doi:10.1103/PhysRevD.99.094023
[arXiv:1811.10911 [hep-ph]].

\bibitem{Zhou:2018bkn}
Q.~S.~Zhou, K.~Chen, X.~Liu, Y.~R.~Liu and S.~L.~Zhu,
Phys. Rev. C \textbf{98} (2018) no.4, 045204
doi:10.1103/PhysRevC.98.045204
[arXiv:1801.04557 [hep-ph]].

\bibitem{Wang:2018lhz}
Z.~G.~Wang,
Eur. Phys. J. C \textbf{78} (2018) no.10, 826
doi:10.1140/epjc/s10052-018-6300-4
[arXiv:1808.09820 [hep-ph]].

\bibitem{Xing:2018bqt}
Y.~Xing and R.~Zhu,
Phys. Rev. D \textbf{98} (2018) no.5, 053005
doi:10.1103/PhysRevD.98.053005
[arXiv:1806.01659 [hep-ph]].

\bibitem{Xing:2019wil}
Y.~Xing,
Eur. Phys. J. C \textbf{80} (2020) no.1, 57
doi:10.1140/epjc/s10052-020-7625-3
[arXiv:1910.11593 [hep-ph]].

\end{references}
  \end{document}